\documentclass[10pt,twocolumn,letterpaper]{article}

\usepackage[pagenumbers]{cvpr} 
\usepackage[dvipsnames]{xcolor}

\definecolor{cvprblue}{rgb}{0.21,0.49,0.74}
\usepackage[pagebackref,breaklinks,colorlinks,citecolor=cvprblue]{hyperref}
\usepackage{graphicx}
\usepackage{amsmath}
\usepackage{amssymb}
\usepackage{booktabs}
\usepackage{CJKutf8}  
\usepackage{array}
\usepackage{tabularx}
\usepackage{multirow}
\usepackage{breakcites} 
\usepackage{comment}
\usepackage{url}
\usepackage{colortbl}
\usepackage{nccmath}
\usepackage{makecell}

\usepackage{times}
\usepackage{epsfig}
\usepackage{mmstyle}

\usepackage[capitalize]{cleveref}
\crefname{section}{Sec.}{Secs.}
\Crefname{section}{Section}{Sections}
\Crefname{table}{Table}{Tables}

\newcolumntype{Y}{>{\centering\arraybackslash}X}
\usepackage{xcolor}
\usepackage{wrapfig}
\usepackage{soul}

\definecolor{yellow}{rgb}{1,1, 0.6}
\definecolor{lightyellow}{rgb}{1,1, 0.8}
\definecolor{orange}{rgb}{1, 0.8, 0.6}
\definecolor{coral}{RGB}{246,131,65}
\definecolor{pinkred}{rgb}{1, 0.6, 0.6}
\definecolor{hotpink}{RGB}{238,64,195}
\definecolor{lavender}{RGB}{207,226,243}
\definecolor{gainsboro}{RGB}{208,224,227}
\definecolor{gainsboro2}{RGB}{217,234,211}
\definecolor{blanchedalmond}{RGB}{252,229,205}

\begin{document}
\begin{CJK}{UTF8}{}
\CJKfamily{mj}

\newcommand{\SU}[1]{{\textcolor{blue}{#1}}}

\def\confName{CVPR}
\def\confYear{2024}

\title{Dispersed Structured Light for Hyperspectral 3D Imaging}

\author{Suhyun Shin$^1$ ~ ~ ~
Seokjun Choi$^1$ ~ ~ ~
Felix Heide$^2$ ~ ~ ~
Seung-Hwan Baek$^1$ \\[2mm]
$^1$ POSTECH ~ ~ ~  ~ ~ ~  $^2$ Princeton University
}


\twocolumn[{
\renewcommand\twocolumn[1][]{#1}
\maketitle
\vspace{-30pt}
\begin{center}
    \centering
    \captionsetup{type=figure}
    \includegraphics[width=175mm]{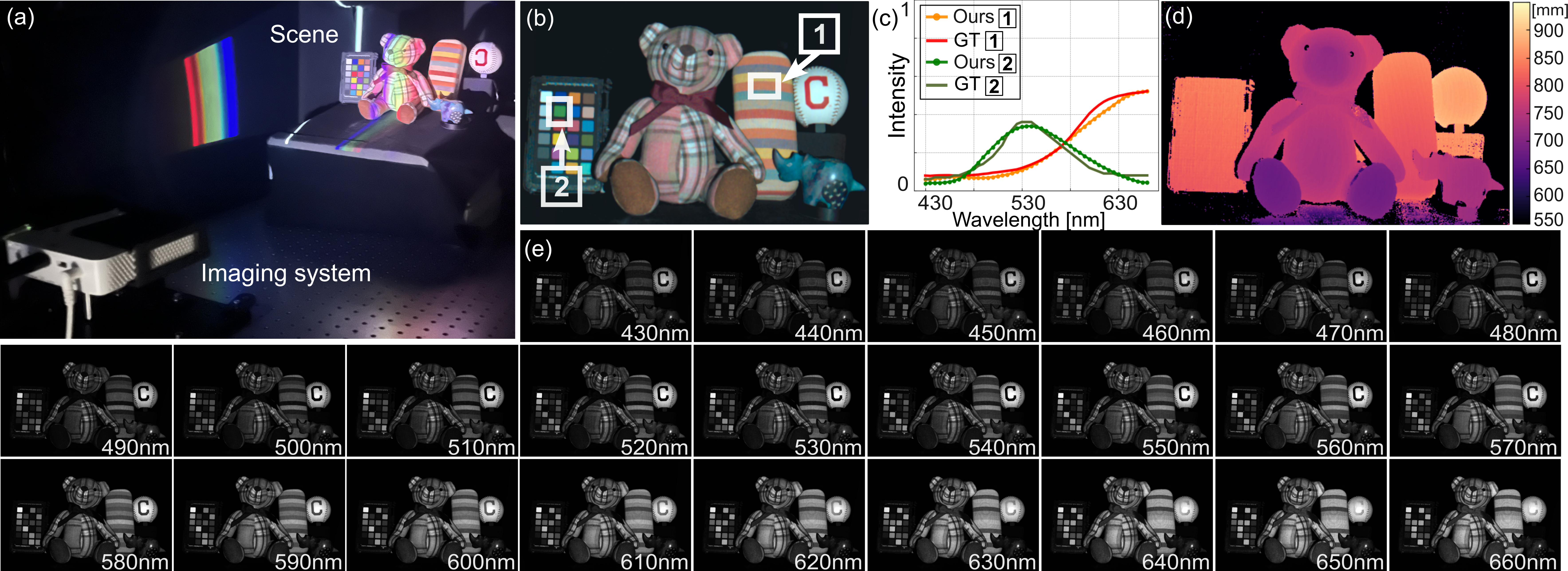}
    \vspace{-20pt}
    \captionof{figure}{ We introduce dispersed structured light, a low-cost high-quality hyperspectral 3D imaging method. By placing a diffraction grating film on a conventional camera-projector setup, we disperse structured-light patterns. Analyzing the images captured under the dispersed structured light enables accurate hyperspectral 3D reconstruction. (a) Capture configuration, (b) estimated hyperspectral image in sRGB, (c) comparison with spectroradiometer measurements, (d) estimated depth map, (e) estimated hyperspectral image.}
    \label{fig:teaser}
\end{center}
}]

\begin{abstract}
Hyperspectral 3D imaging aims to acquire both depth and spectral information of a scene. However, existing methods are either prohibitively expensive and bulky or compromise on spectral and depth accuracy. In this paper, we present Dispersed Structured Light (DSL), a cost-effective and compact method for accurate hyperspectral 3D imaging. DSL modifies a traditional projector-camera system by placing a sub-millimeter thick diffraction grating film front of the projector.
This configuration enables dispersing structured light based on light wavelength. 
To utilize the dispersed structured light, we devise a model for dispersive projection image formation and a per-pixel hyperspectral 3D reconstruction method. 
We validate DSL by instantiating a compact experimental prototype.
DSL achieves spectral accuracy of 18.8\,nm full-width half-maximum (FWHM) and depth error of 1\,mm, outperforming prior work on practical hyperspectral 3D imaging.
DSL promises accurate and practical hyperspectral 3D imaging for diverse application domains, including computer vision and graphics, cultural heritage, geology, and biology.
\end{abstract}

\section{Introduction}
Hyperspectral 3D imaging aims to capture both depth and spectrum per pixel. Allowing for geometric-spectral analysis of a scene~\cite{baek2021single,li2022deep,li2021spectral,li2019pro,wang2016simultaneous}, this imaging modality has potential applications across diverse domains, including food ripeness detection~\cite{sun2010hyperspectral}, mineral detection~\cite{TRIPATHI2019e02931}, art authentication~\cite{polak2017hyperspectral}, classification~\cite{mahesh2015hyperspectral}, and cultural heritage preservation~\cite{kim20123d}. 

Despite of the potential in capturing rich geometric and spectral information, existing hyperspectral 3D imaging methods are often impractical, due to the high instrumentation costs, large form factor, and low accuracy. 
Specifically, high-end systems such as using coded-aperture snapshot spectral imaging systems (CASSI) and liquid-crystal tunable filters provide accurate depth and spectral information, however mandate high cost and large form factor~\cite{heist20185d,kim20123d, diaz2018hyperspectral+, rueda2019snapshot, wang2016simultaneous, xiong2017snapshot, zhu2018hyperspectral, xu2020hyperspectral}. 
In contrast, affordable hyperspectral 3D imaging systems suffer from low accuracy of depth and spectrum~\cite{li2022deep, kitahara2015simultaneous, hirai2016measuring, li2019pro, li2021spectral, nam2014multispectral, ozawa2017hyperspectral, baek2021single}.

In this paper, we exploit dispersion, a phenomenon where light rays are spatially redirected according to their wavelength either by refraction or diffraction.
Using the optical dispersion, we present dispersed structured light aiming for practical and accurate hyperspectral 3D imaging.
DSL augments a conventional projector-camera system with diffraction grating film, which adds a negligible cost and size of around 10 USD with a sub-millimeter thickness. The diffraction grating in front of the projector spatially disperses broadband projector light based on its wavelength, resulting in a dispersed pattern consisting of several diffraction orders~\cite{Hecht:2001:Optics}. 

Specifically, zero-order diffraction permits light to pass through the diffraction grating as if the grating is not present, enabling us to exploit it for 3D imaging. 
First-order diffraction changes the direction of a light ray based on wavelength, resulting in dispersed structured light patterns that aid hyperspectral reconstruction. 
We develop a model for dispersive projection image formation and a per-pixel hyperspectral 3D reconstruction method using multi-order diffractions. 

DSL achieves an average depth error of {1\,mm} and spectral FWHM of 18.8\,nm in the visible spectrum, outperforming existing practical hyperspectral 3D imaging method~\cite{li2019pro} that has a spectral FWHM of 83\,nm. 

In summary, we make the following contributions.
\begin{itemize}
\item We present dispersed structured light that enables compact, low-cost, and high-quality hyperspectral 3D imaging by augmenting a projector-camera setup with a diffraction grating film.
\item We develop a model for disperisve projection image formation and per-pixel hyperspectral 3D reconstruction method that considers both zero-order and first-order diffractions. 
\item We perform extensive evaluations and show that DSL achieves an average depth error of 1\,mm and spectral FWHM of 18.8\,nm, outperforming the state-of-the-art affordable hyperspectral 3D imaging methods.
\end{itemize}
\section{Related Work}
\paragraph{Hyperspectral 3D Imaging} 
{Existing work on hyperspectral 3D imaging typically combines the field of hyperspectral imaging~\cite{makarenko2022real, li2022quantization, cai2022coarse,zhang2022herosnet, li2023pixel} with depth imaging~\cite{gigadepth, wang2016simultaneous, hirai2016measuring}.}
Specifically, researchers have paired CASSI~\cite{wagadarikar2008single} with structured light~\cite{diaz2018hyperspectral+,kim20123d}, time-of-flight camera~\cite{rueda2019snapshot}, light-field camera~\cite{xiong2017snapshot}, and stereo~\cite{wang2016simultaneous}.
An alternative to CASSI is the use of spectral bandpass filters with structured light~\cite{heist20185d,baek2021single} or a light field camera~\cite{zhu2018hyperspectral}. 
Xu et al.\cite{xu2020hyperspectral} present a hyperspectral projector that incorporates collimation optics, diffraction grating, and digital micro-mirror device.
Although these systems acquire accurate hyperspectral 3D information, their increased complexity due to relay lenses, narrow-band spectral filters, and plate-based dispersion optics results in a high building cost and large form factor.
 
Several methods have explored hyperspectral 3D imaging with compact setups.
Baek et al.~\cite{baek2021single} propose a diffractive optical element (DOE) producing a point spread function (PSF) that varies with scene spectrum and depth, facilitating single-shot hyperspectral 3D imaging through blur analysis. 
However, the depth and spectral reconstruction accuracy are limited due to the low-frequency characteristics of the PSF.
Li et al.~\cite{li2022deep,li2019pro} employ a projector-camera setup to capture a scene with varying trichromatic projector primaries with known spectra, allowing for hyperspectral reconstruction. 
Despite its practicality, the spectral accuracy is limited by large spectral bandwidth of the projector primary spectra, resulting in {83\,nm} FWHM.
In comparison, the proposed DSL enables accurate hyperspectral 3D imaging with {18.8\,nm} FWHM and average depth error of 1\,mm by dispersing projector light at a cost and form factor on-par with conventional structured light systems.

\paragraph{Structured Light}
Structured light techniques project illumination patterns onto a scene and analyze the reflected light using a camera~\cite{Geng:11}.
The projected patterns enable establishing correspondence between camera and projector pixels for 3D imaging.
DOEs, combined with narrow-band coherent laser, have often been employed as cost-effective components for generating structured light~\cite{baek2021polka, kim2022metasurface, ni2020metasurface}.
Employing a conventional projector instead of the DOE-based laser illumination allows for using multiple patterns in a programmable manner, significantly enhancing depth accuracy~\cite{geng2011structured,gigadepth}.
Various structured light patterns have been proposed to facilitate 3D imaging robust to global illumination~\cite{nayar2006fast, gupta2013structured}, light transport analysis~\cite{gupta2009focusing, o20143d}, and energy-efficient 3D imaging~\cite{o2015homogeneous}.
Introducing polarizing optics to a conventional projector further enhances its capability, making 3D imaging of translucent objects and polarimetric light transport analysis feasible~\cite{baek2021polarimetric, jeon2023polarimetric}.
The proposed DSL augments structured light by only placing a diffraction grating film in front of a projector, enabling accurate hyperspectral 3D imaging with a compact setup.

\paragraph{Dispersive Optics for Cameras and Projectors}
Dispersive optics, such as prism and diffraction grating, have been often employed in the design of cameras and displays, in particular for hyperspectral imaging.
CASSI employs dispersive optics and a coded mask to obtain masked spectral images with a wavelength-dependent translation~\cite{wagadarikar2008single}.
Using a prism and a coded mask without relay lenses enables video-rate hyperspectral imaging~\cite{cao2011high}. Computed tomography imaging spectrometers leverage multi-order dispersion from a diffraction grating for hyperspectral imaging~\cite{Narea-Jimenez:22}. Hostettler et al.\cite{hostettler15dispersion} use a prism and a mask to implement a trichromatic color projector based on dispersion. Mohan et al.~\cite{mohan2008agile} employ a diffraction grating and an attenuation mask to control the spectral power distribution of projector light. 
Recently, Sheinin et al.\cite{sheinin2020diffraction,sheinin2021deconvolving} use a diffraction grating and a line camera to track fast-moving sparse scene points. 
Our DSL augments a projector-camera system only with a diffraction grating film, allowing to keep the form factor compact and cost low. Instead of directly masking the spectrum emitted from the projector using additional components, we analyze and exploit how structured light patterns are dispersed and illuminate a scene.
\section{Dispersive Projection Image Formation}
\label{sec:image_formation}

\paragraph{Imaging Setup}
We devise the proposed DSL imaging system with the goals of compactness and affordability. To this end, we combine a conventional trichromatic camera (FLIR GS3-U3-32S4C-C) with a trichromatic projector (LG PH30N) and a diffraction-grating film (Edmund \#54-509) mounted in front. This configuration makes light from the projector undergo dispersion, and, as such, patterns emitted from the projector are spatially dispersed depending on wavelength. Figure~\ref{fig:system} shows our experimental prototype and a captured image with a projector pattern composed of four squares. For zero-order diffraction, light does not experience wavelength-dependent changes, and the projected squares retain their original shapes. For first-order diffractions, each square undergoes wavelength-dependent shift.

\subsection{Image Formation}
\paragraph{Background on Diffraction Grating} 
Diffraction grating consists of repetitive, even-spaced micron-scale grooves characterized by the groove density $g$. 
The interaction of light with the grating incurs wavelength-dependent diffraction~\cite{Hecht:2001:Optics}, which is modeled in the geometric-optics perspective as a redirection of the light path:
\begin{align}
\texttt{dg}(\mathbf{v}, m, \lambda)= (\underbrace{{-m g \lambda} + {v}_x}_{d_x}, \underbrace{{v}_y}_{d_y}, \sqrt{1 - {d}^2_x - {d}^2_y}),
\label{eq:grating}
\end{align}
where $m$ is the diffraction order, $\mathbf{v}=(v_x,v_y,v_z)$ is the incident light vector. 
$x$-axis here is aligned with the groove direction and $\lambda$ is the wavelength.
$\texttt{dg}(\mathbf{v}, m, \lambda)$ is the ray direction after diffraction. 
In addition to the direction change, the diffraction grating introduces a change in light intensity depending on diffraction order and wavelength, represented as diffraction efficiency $\eta_{m,\lambda}$~\cite{Hecht:2001:Optics}. 

\begin{figure}[t]
	\centering
		\includegraphics[width=\columnwidth]{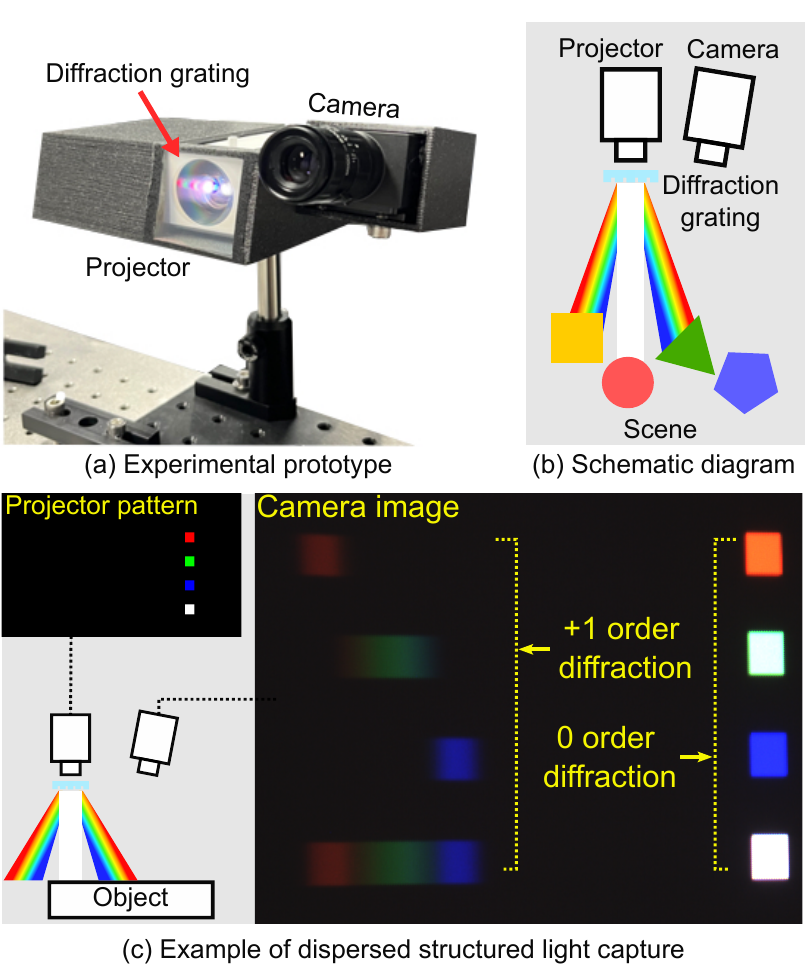}
		\caption{ 
  \textbf{Experimental prototype.} (a) \& (b) Our prototype consists of an RGB projector equipped with a diffraction grating film, and an RGB camera. (c) An example projector pattern and its corresponding captured image, exhibiting clear first-order diffraction.
}
\vspace{-5mm}
		\label{fig:system}
\end{figure}

We use a diffraction grating which creates positive and negative first-order diffractions appearing to the left and right sides with respect to the zero-order component as shown in Figure~\ref{fig:system}(c).
Note that higher-order diffractions are not detected inside of our camera FoV.

%

\begin{figure*}[t]
	\centering
		\includegraphics[width=\linewidth]{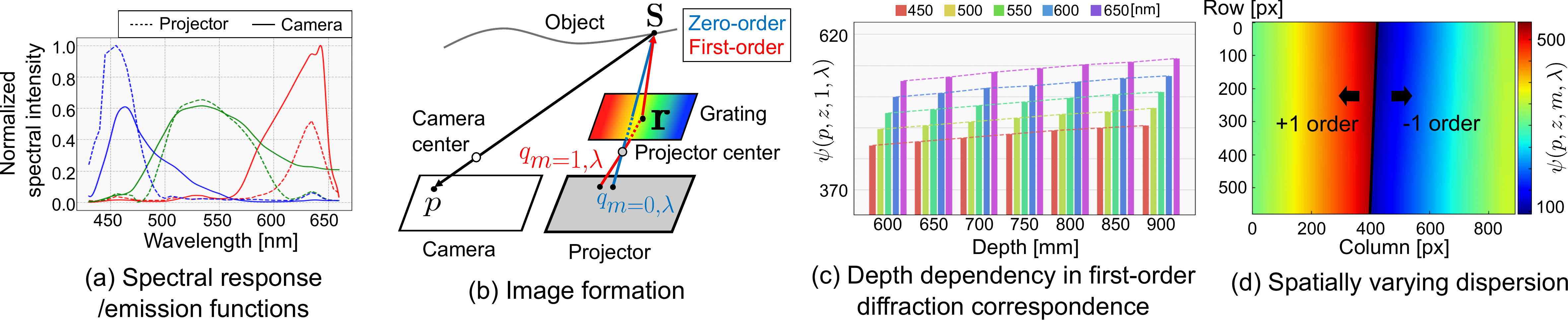}
  \caption{\textbf{Image formation.} (a) Camera response function and projector emission function. (b) Schematic diagram of image formation. (c) Depth dependency of the correspondence function $\psi$ for $m=1$ and a camera pixel $p$. (d) Spatially-varying correspondence map for depth 700\,mm and wavelength 430\,nm for the first-order diffractions $m=-1,1$.
  }
  \vspace{-3mm}
		\label{fig:image_function}
\end{figure*}
\paragraph{Forward Model}
For a projector pattern $P(q,c)$, where $q$ is a projector pixel and $c$ is a color channel (R, G, or B), the light intensity $L(q,\lambda)$ emitted for each wavelength $\lambda$ is 
\begin{equation}
L(q, \lambda) =  \sum_{c} \Omega^{\text{proj}}_{c, \lambda} P(q,c),
\label{eq:DSL}
\end{equation}
where $\Omega_{c,\lambda}^{\text{proj}}$ is the projector emission function shown in Figure \ref{fig:image_function}(a).

The light then passes through the diffraction grating and splits into multiple light rays with varying propagation directions $\texttt{dg}(\mathbf{v}, m, \lambda)$, depending on diffraction order $m$ and wavelength $\lambda$. 
Here, we focus on a ray corresponding to wavelength $\lambda$ and diffraction order $m$.
The ray will propagate and eventually illuminate a scene point {$\mathbf{S}$} at depth $z$.
The reflected light from the scene point will be captured by a camera pixel $p$. 
We then establish a correspondence function $\psi$ that maps the camera pixel $p$, depth $z$, diffraction order $m$, and wavelength $\lambda$ to the corresponding projector pixel $q_{m,\lambda}$ that emitted the light ray illuminating the scene point:
\begin{equation}
    \label{eq:correspondence}
    q_{m,\lambda} = \psi\left(p, z, m,\lambda\right).
\end{equation}
We proceed to describe our full image formation before describing how we model the correspondence function $\psi$.

The intensity at the camera pixel $p$ and color channel $c$ then can be modeled as 
\begin{align}
I(p,c) = \sum_{\lambda \in \Lambda} \Omega^{\text{cam}}_{c, \lambda}  H(p,\lambda) \underbrace{\frac{1}{d(p)^2} \sum_{m=-1}^{1} \eta_{m,\lambda} L(q_{m,\lambda}, \lambda) }_{\text{Dispersed structured light}},
\label{eq:image_formation}
\end{align}
where $\Omega_{c,\lambda}^{\text{cam}}$ is the camera response function shown in Figure \ref{fig:image_function}(a), $\Lambda=[\lambda_1,\cdots,\lambda_{47}]$ is the set of 47 wavelengths sampled from 430\,nm to 660\,nm with 5\,nm interval. $H(p,\lambda)$ is the hyperspectral intensity, which we aim to estimate. 
$d(p)$ is the propagation distance from the projector to the scene point {$\mathbf{S}$}, as such, the term $1/d(p)^2$ describes inverse-square law. 

The dispersed structured light term in Equation~\eqref{eq:image_formation} describes the light intensity projected to the scene point {$\mathbf{S}$} by aggregating zero- and first-order light energy $L(q_{m,\lambda}, \lambda)$ weighted by the corresponding diffraction efficiency $\eta_{m,\lambda}$.
In the following, we model the correspondence function $\psi$ introduced in Equation~\eqref{eq:correspondence}.

\subsection{Correspondence under Dispersion}
\label{sec:correspondence}

\paragraph{Zero-order Diffraction}
For zero-order diffracted light ($m=0$), light transport can be analyzed as if no diffraction grating exists~\cite{Hecht:2001:Optics}: given a camera pixel $p$ and its depth $z$, we obtain the corresponding projector pixel $q_{0,\lambda}$ for any visible wavelength $\lambda$ by applying perspective unprojection and projection~\cite{hartley2003multiple}:
\begin{align}
    \label{eq:correspondence_zero}
    q_{0, \lambda} &= \psi\left(p, z,0,  \lambda\right) \\
    &= \texttt{project}\left(\texttt{unproject}\left(p,z\right)\right),
\end{align}
where $\texttt{unproject}(\cdot)$ is the perspective unprojection from a camera pixel $p$ to a 3D point {$\mathbf{S}$} using the depth $z$. 
$\texttt{project}(\cdot)$ is the perspective projection from the 3D point {$\mathbf{S}$} to the projector.

\vspace{-2mm}
\paragraph{First-order Diffractions}
Correspondence for first-order diffractions ($m=-1 \text{ or } 1$) is more challenging to model. Due to the direction change by diffraction grating, direct perspective projection from the scene point {$\mathbf{S}$} to the projector is no longer valid. Instead, we need to identify a point $\mathbf{r}$ on the diffraction grating where the light ray from the projector center $\mathbf{t}_\text{proj}$ diffracts at $\mathbf{r}$ and reaches to the scene point {$\mathbf{S}$}. This is illustrated in Figure~\ref{fig:image_function}(b) and by using Equation~\eqref{eq:grating} it can be formulated as
\begin{align}
\underset{\mathbf{r}}{\text{minimize}} \, \| \texttt{dg}({\dot{\overrightarrow{\mathbf{t}_\text{proj}\mathbf{r}}}} , m, \lambda ) - {\dot{\overrightarrow{\mathbf{r} {\mathbf{S}} }}} \|_2,
\label{eq:projection_equality}
\end{align}
where $\dot{\overrightarrow{x}}$ represents the normalized vector of $\overrightarrow{x}$. 
While the formulation is intuitive, solving Equation~\eqref{eq:projection_equality} over the entire 3D volume, for each wavelength $\lambda$ and diffraction order $m$, is prone to calibration errors and is computationally demanding when approached with iterative numerical methods like the Newton-Raphson method or root-finding algorithms.

Hence, instead of directly solving Equation~\eqref{eq:projection_equality}, we develop a data-driven model for efficient and accurate correspondence mapping of the first-order diffractions. 
We prepare samples: first-order corresponding projector pixel positions $q'_{m,\lambda}$ for wavelengths $\lambda' \in \mathbb{S}_\lambda$, grid-sampled camera pixels $p'\in\mathbb{S}_p$, and depths $z'\in \mathbb{S}_z$.
Section~\ref{sec:calibration} describes the sample acquisition.
We then fit an exponential function to the samples for modeling depth dependency of the correspondence function $\psi$ and perform linear interpolation for spatial and spectral dimensions.
As the first-order diffraction in our prototype occurs along the horizontal axis, and our hyperspectral reconstruction only uses horizontal correspondence, we only model the horizontal coordinate of the correspondence function, i.e., the column index.
More details on our data-driven correspondence model can be found in the Supplemental Document.

Our data-driven first-order correspondence model obtains a mean reprojection error of 1px and can be efficiently evaluated with tabulation in $O(1)$.
Figure~\ref{fig:image_function}(c) shows the depth dependency of the modeled correspondence function $\psi$. Figure \ref{fig:image_function}(d) shows the spatially-varying correspondence map for depth $700$\,mm, wavelength $430$\,nm, and the diffraction orders $m=1 \text{ and } -1$.
\vspace{-2mm}
\section{Hyperspectral 3D Reconstruction}
\label{sec:reconstruction}
With the image formation model from above in hand, we estimate a depth map using binary-code structured light and then {reconstruct the hyperspectral image using scanline structured light.}

\subsection{Depth Reconstruction} 

\begin{figure}[t]
	\centering
		\includegraphics[width=\columnwidth]{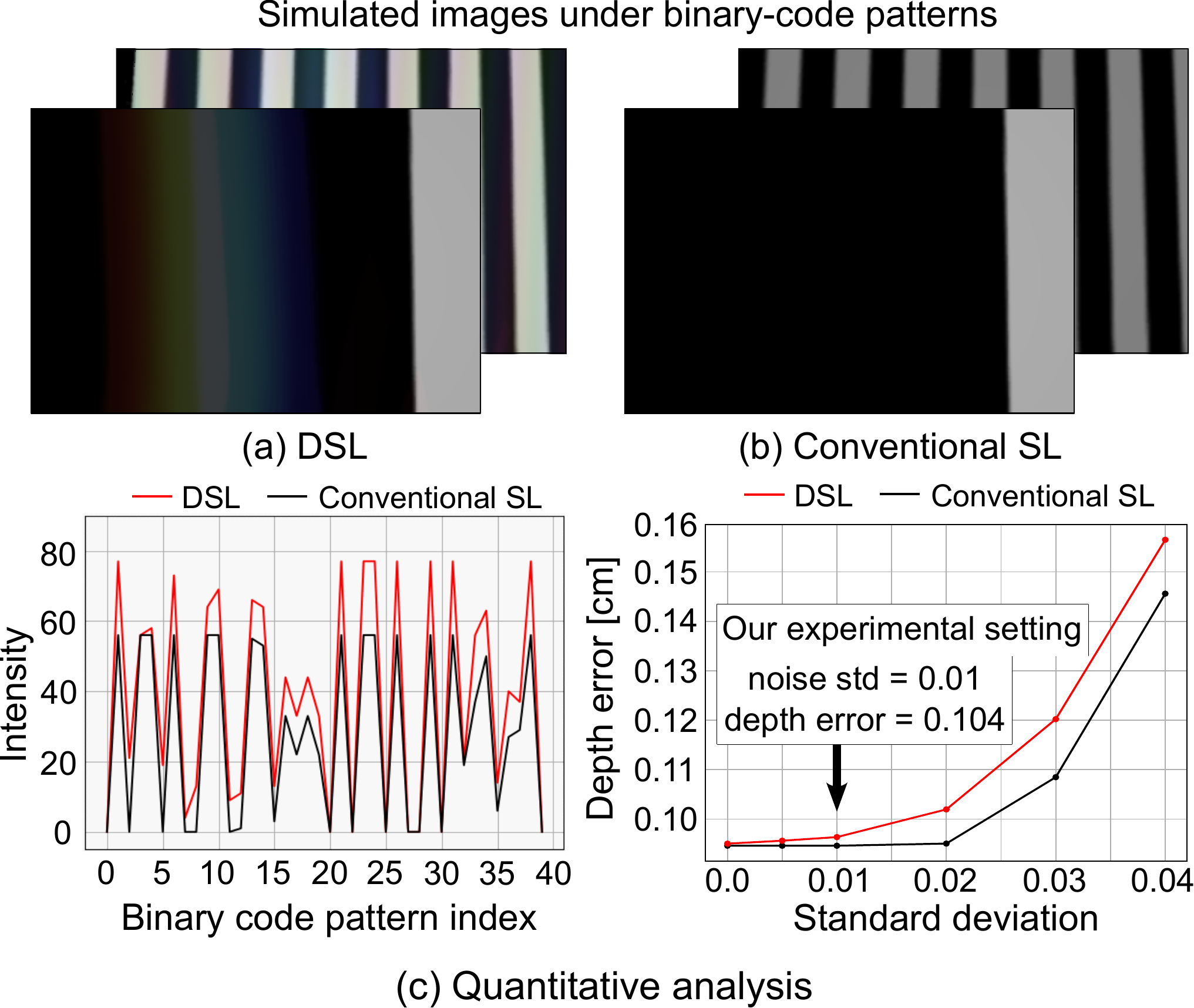}
		\caption{ \textbf{Binary decoding under dispersion.} Simulated images under the binary-code patterns for (a) DSL and (b) conventional SL. 
  (c) Intensity of a camera pixel with DSL and conventional SL.
  (d) Depth error of binary-code decoding for DSL and conventional SL with varying Gaussian noise.
  } 
  \vspace{-3mm}
		\label{fig:depth_reconstruction}
\end{figure}

For depth estimation, we use binary-code structured light patterns shown in Figure~\ref{fig:depth_reconstruction} and represented as $[P_1,\dots, P_{K_b}]$: $P_i(q,\forall c) = \texttt{bit}(q,i)$, where $\texttt{bit}(q,i)$ returns the $i$-th binary bit of the decimal horizontal and vertical location of the projector pixel $q$, encoding the projector pixel location as a binarized code~\cite{geng2011structured}.
$K_b=40$ is the number of patterns.
Then, each camera pixel $p$ observes the fluctuating intensity values under these binary-code patterns, producing an observation vector $[I_1(p,\forall c), \dots, I_{K_b}(p,\forall c)]$ shown in Figure~\ref{fig:depth_reconstruction}(c).

\paragraph{Binary-code Decoding}
From the observation vector, we aim to estimate the zero-order correspondence $q_{0,\lambda}$, which will facilitate the estimation of depth $z$ through triangulation using the zero-order correspondence function $\psi(p,z,0,\lambda)$.
Note that zero-order correspondence has no dependency on wavelength $\lambda$.
For the $i$-th projector pattern $P_i$, the conventional decoding method binarizes the captured intensity $I_i$ into 0 or 1 based on its intensity level by using RGB-to-gray conversion and thresholding with a constant $\tau$.
The binary code is then decoded to a decimal number, which is the location of the corresponding projector pixel $q_{0,\lambda}$. 

\paragraph{Validation of Binary-code Decoding for DSL}
An important question is whether the conventional binary decoding works for DSL in the presence of dispersed light.
We provide both mathematical and experimental verification.
In theory, this can be confirmed if the intensity lit by the zero-order diffracted light exceed the intensity when not receiving the zero-order diffracted light. 
We derive the minimum captured intensity $I^\text{on}_\text{min}$ when a scene point is only illuminated by the zero-order structured light.
Also, we model the maximum captured intensity $I^\text{off}_\text{max}$ when the zero-order structured light does not illuminate the point and the first-order light illuminates.
The quantities are written as 

\begin{align}
I^\text{on}_\text{min}(p,c)&=\sum_{\lambda } \Omega^{\text{cam}}_{c, \lambda} H(p,\lambda) \eta_{0,\lambda} \sum_{c'} \Omega^{\text{proj}}_{c', \lambda},\\
I^\text{off}_\text{max}(p,c) &= \sum_{\lambda } \Omega^{\text{cam}}_{c, \lambda} H(p,\lambda) \sum_{m={1,-1}} \eta_{m,\lambda} \sum_{c'} \Omega^{\text{proj}}_{c', \lambda}.
\label{eq:intensity_min_max}
\end{align}

We then compare both quantities, giving an observation that the inequality $I^\text{off}_\text{max} < I^\text{on}_\text{min}$ holds in our setup configuration, because for all wavelength $\lambda$ the following inequality holds $\eta_{0,\lambda} \sum_{c'} \Omega^{\text{proj}}_{c', \lambda} > \sum_{m={1,-1}} \eta_{m,\lambda} \sum_{c'} \Omega^{\text{proj}}_{c', \lambda}$.
This validates the use of conventional binary-code decoding for DSL in theory. Refer to the Supplemental Document for the derivation details.

Next, we further validate the applicability of binary decoding on {$10^6$} simulated samples with random hyperspectral reflectance for a planar object with Gaussian measurement noise. Example observation vectors with and without considering first-order diffractions are shown in Figure~\ref{fig:depth_reconstruction}(c). We set the Gaussian standard deviation $0.01$ corresponding to the real-world noise in our hardware measurements. The average depth error after decoding, shown in Figure~\ref{fig:depth_reconstruction}(d), is $1.04$\,mm in the presence of first-order diffractions. 
This experiment further validates the use of binary-code decoding for DSL. 

\subsection{Hyperspectral Reconstruction}
\label{sec:hyperspectral_reconstruction}
Once depth $z$ is obtained, we proceed to estimate per-pixel {hyperspectral intensity $\mathbf{H} =[H(p,\lambda_1), \cdots, H(p,\lambda_N)] \in \mathbb{R}^{N\times1}$.} To this end, we use the scanline structured light patterns $[P_1,,..., P_{K_s}]$ that scans through the whole column with a line width $w$. 
$K_s=318$ is the number of scanline patterns. Figure~\ref{fig:hyper_reconstruction} shows that the first-order diffracted light from the $i$-th scanline pattern produces a narrow-band illumination spread across multiple columns. Consequently, scanning columns from left to right enables illuminating each scene point with every narrow-band light, facilitating high-quality hyperspectral reconstruction.

\paragraph{Pixel Intensity under Scanline Patterns}
For the $i$-th scanline pattern, a camera pixel $p$ receives either zero-order diffraction, first-order diffractions, or no illumination at all. This simplifies our image formation as 
\begin{align}
    I_i(p,c) = 
    \begin{cases}
      \sum_{\lambda} \Omega^{\text{cam}}_{c, \lambda}  H(p,\lambda) \eta_{0,\lambda} L(q_{0,\lambda}, \lambda) & \text{for zero order},\\
      \Omega^{\text{cam}}_{c, \lambda}  H(p,\lambda) \eta_{m,\lambda} L(q_{m,\lambda}, \lambda) & \text{for first orders},\\
      0 & \text{otherwise}.
    \end{cases}       
    \label{eq:image_formation_white_line} 
\end{align}
Figure~\ref{fig:hyper_reconstruction}(c) shows the intensity graph of a camera pixel $p$ for varying index of the scanline patterns: $[I_1(p,c), \cdots, I_{K_s}(p,c)]$.
For the first-order cases ($m=-1\text{ or }1$), the intensity stems from a specific wavelength $\lambda$, hence narrow-band illumination. 

\paragraph{Index Mapping for First-order Diffraction}
{
We exploit first-order diffractions for accurate  hyperspectral reconstruction.
We find the scanline pattern index $i$ of which $m$-order diffracted light with wavelength $\lambda$ illuminates the camera pixel $p$.
In fact, we already obtained this index mapping in the form of the correspondence function $\psi$: 
\begin{equation}
\delta_{m,i} = \texttt{px2index}\left(\psi\left(p,z,m,\lambda_i\right)\right),
    \label{eq:correspondence_m}
\end{equation}
where $\delta_{m,i}$ is the corresponding scanline pattern index and $\texttt{px2index}$ is the conversion function that simply returns the scanline pattern index that lights up at the input projector pixel location.

We then collect the intensity captured under narrow-band illumination for all target wavelengths in $[\lambda_1, \cdots, \lambda_N]$ for each diffraction order $m$: 
\begin{equation}
    \mathbf{I}_{m} =[ I_{\delta_{m,1}} (p,c), \cdots, I_{\delta_{m,N}} (p,c) ] \in \mathbb{R}^{N \times 3}.
    \label{eq:I_hyp}
\end{equation}

\begin{figure}[t]
	\centering
		\includegraphics[width=\columnwidth]{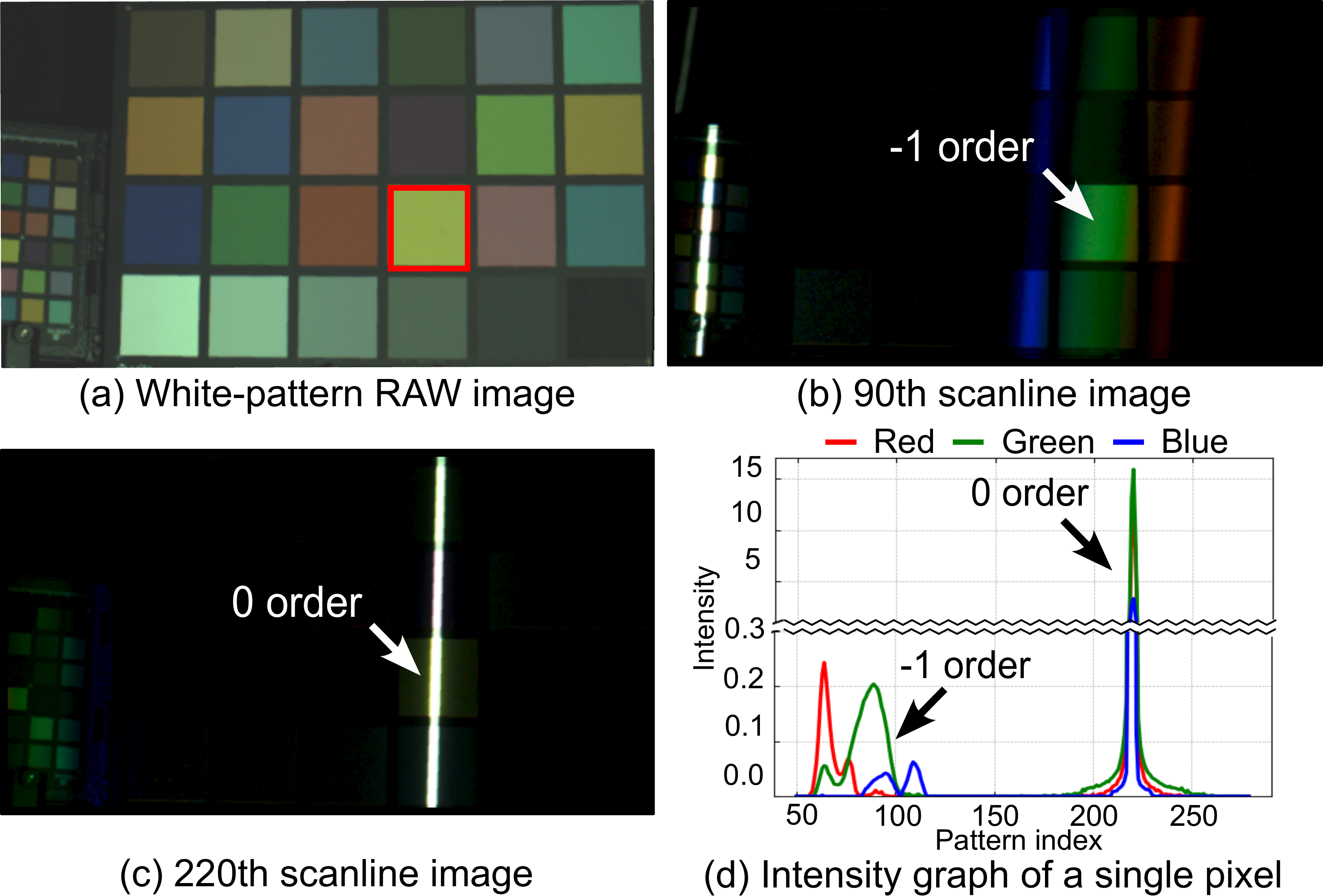}
		\caption{ 
  \textbf{Hyperspectral imaging with scanline illuminations.}
  Captured images under (a) white pattern and (b)\&(c) scanline patterns. Narrow-band illumination over multiple columns is shown in (b), which originates from the first-order diffraction.  (d) Pixel intensity with respect to varying scanline pattern index.}
  \vspace{-4mm}
		\label{fig:hyper_reconstruction}
\end{figure}

\paragraph{Optimization}
We formulate a per-pixel optimization problem for accurate hyperspectral reconstruction:
\begin{equation}
     \underset{\mathbf{H}}{\text{argmin}} \, 
\underbrace{\sum_{m=-1}^1 \kappa_{m} \| \mathbf{A}_{m}\mathbf{H} - \mathbf{I}_{m}  \|_2^2}_{\text{Data term}}  +  \underbrace{\kappa_\lambda \| \nabla_\lambda \mathbf{H}\|_2^2}_{\text{Regularization term}},
    \label{eq:hyperspectral_optimization2} 
\end{equation}
where $\kappa_m$ is the spatially-varying per-pixel balancing weight for the diffraction order $m$.
{Refer to the Supplemental Document how we set the balancing weight.}
$\mathbf{A}_m$ is the system matrix defined as 
\begin{equation}
    \mathbf{A}_m =
        \begin{cases}
       \sum_\lambda \Omega^{\text{cam}}_{c, \lambda}  \eta_{m,\lambda} L(q_{m,\lambda}, \lambda) & \text{for zero order},\\
      \Omega^{\text{cam}}_{c, \lambda}  \eta_{m,\lambda} L(q_{m,\lambda}, \lambda) & \text{for first orders}.
    \end{cases}       
    \label{eq:element}
\end{equation} 
$\nabla_\lambda$ is the gradient operator along the spectral axis. The first term in Equation~\eqref{eq:hyperspectral_optimization2} accounts for the reconstruction loss across the multiple diffraction orders. The second term is the spectral smoothness prior~\cite{baek2017compact}. We use gradient descent for the optimization, which takes three minutes to converge. Our hyperspectral reconstruction method operates on a per-pixel basis and exploits first-order diffractions, enabling accurate hyperspectral image reconstruction.
}
\section{Calibration}
\label{sec:calibration}
\noindent
We calibrate the image formation parameters of the projector, camera, and diffraction grating, as briefly described in the following. A detailed description of the calibration procedure can be found in the Supplemental Document.

\paragraph{Diffraction Efficiency}
To calibrate the diffraction efficiency $\eta_{m, \lambda}$, we measure the intensity of $m$-order diffracted light at each wavelength $\lambda$ projected onto a Spectralon sample. We use spectral bandpass filters at $10$~nm intervals, from $430$~nm to $660$~nm. Diffraction efficiency is then computed as the intensity ratio of each first-order wavelength measurement over the zero-order intensity. 

\paragraph{Spectral Response and Emission Functions}
The projector spectral emission function, $\Omega^{\text{proj}}_{\lambda,c}$, was obtained by projecting red, green, and blue dots onto a Spectralon target, measuring the reflected radiance with a spectroradiometer (JETI Specbos 1211), and normalizing the results with Spectralon reflectance. For the camera response function, $\Omega^{\text{cam}}_{\lambda,c}$, we use the data provided by the camera manufacturer. For both emission and response functions, we perform refinements of which details can be found in the Supplemental Document. Figure~\ref{fig:image_function}(a) shows the calibrated functions.

\paragraph{First-order Correspondence Model}
To calibrate the first-order correspondence model described in Section~\ref{sec:correspondence}, we acquired images of flat Spectralon surfaces at five different depth positions with scanline illumination patterns present. These images were captured using multiple bandpass filters. For sampled camera pixels, denoted $p'$, we identified the corresponding projector pixel $q'_{m,\lambda}$ from the captured images.
Using the samples, we obtain the data-driven corresponding model $\psi$. 
\section{Assessment}
\label{sec:results}

\begin{figure}[t]
\centering
\includegraphics[width=\columnwidth]{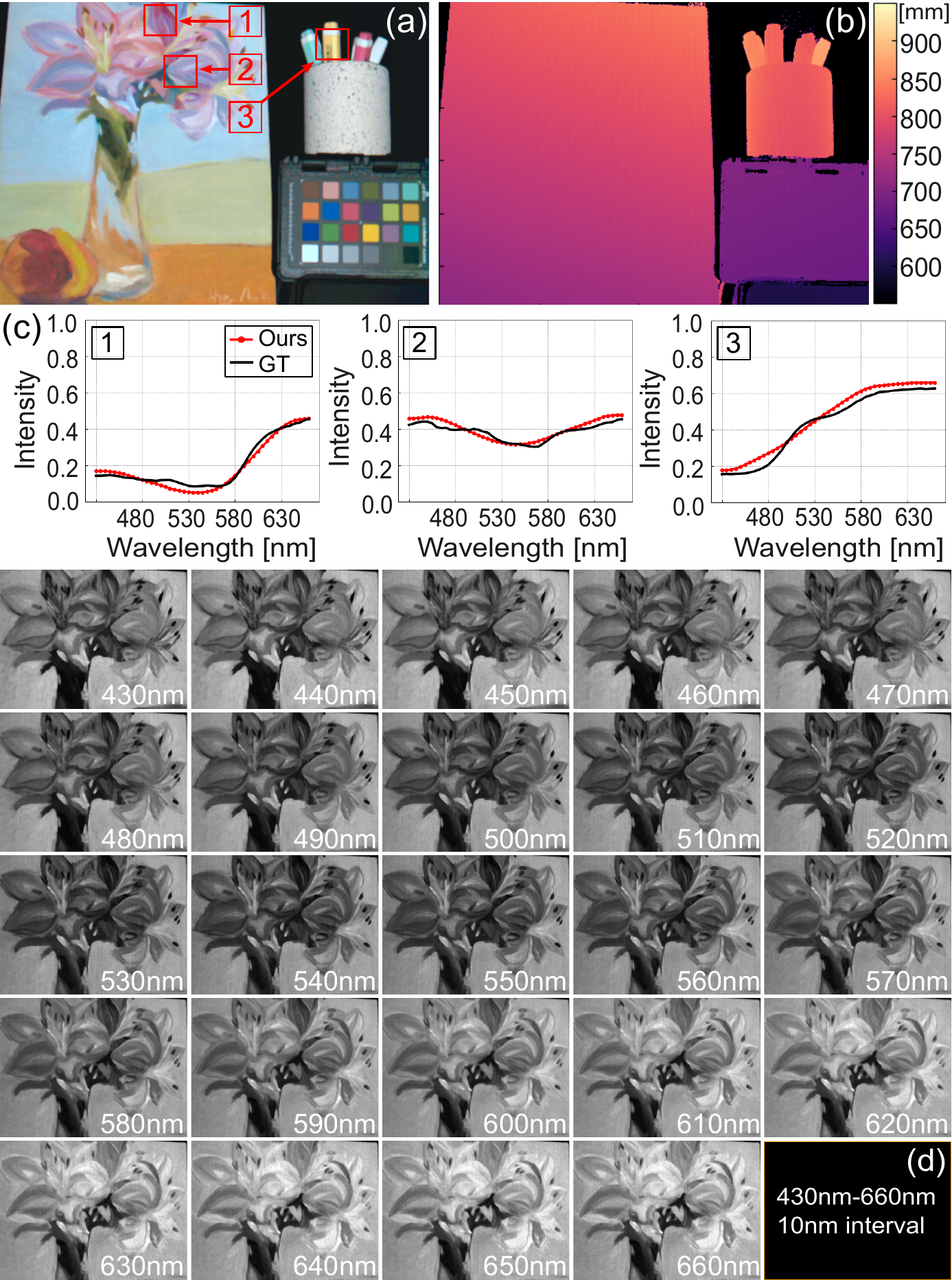}
\caption{\textbf{Hyperspectral 3D imaging.} (a) Reconstructed hyperspectral image visualized in sRGB, (b) reconstructed depth map, (c) estimated hyperspectral intensity for three different points compared with spectroradiometer measurements, (d) estimated hyperspectral image.}
\vspace{-5mm}
\label{fig:extra_scene}
\end{figure}

\begin{figure}[t]
\centering
\includegraphics[width=\columnwidth]{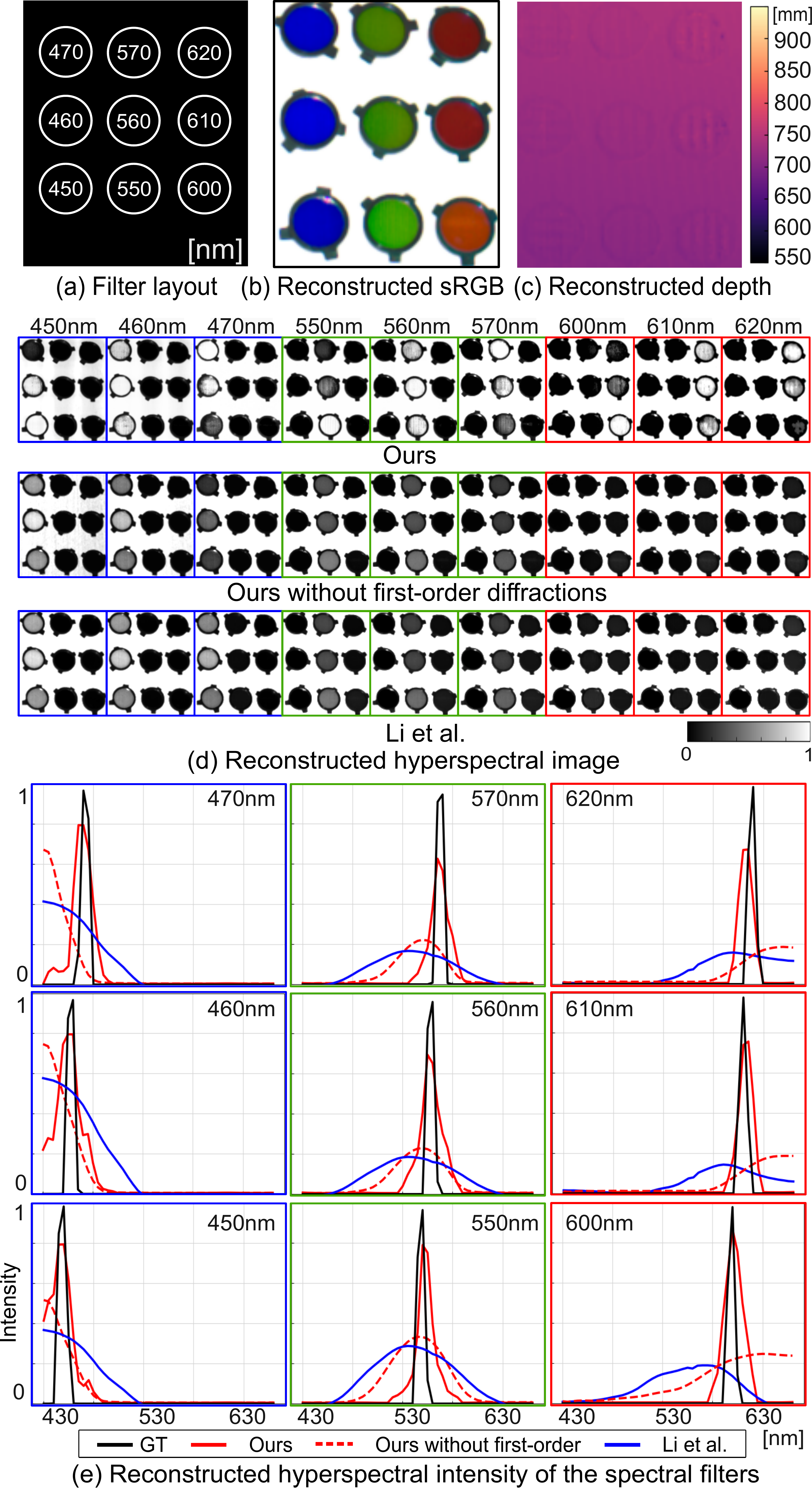}
\caption{\textbf{High-frequency spectral imaging.} (a) Filter layout, (b) reconstructed hyperspectral image in sRGB, (c) reconstructed depth, (d) reconstructed hyperspectral images, (e) spectral curves for the nine bandpass filters. 
}
\vspace{-5mm}

\label{fig:peak_comparison}
\end{figure}

\begin{figure}[t]
\centering
\includegraphics[width=\columnwidth]{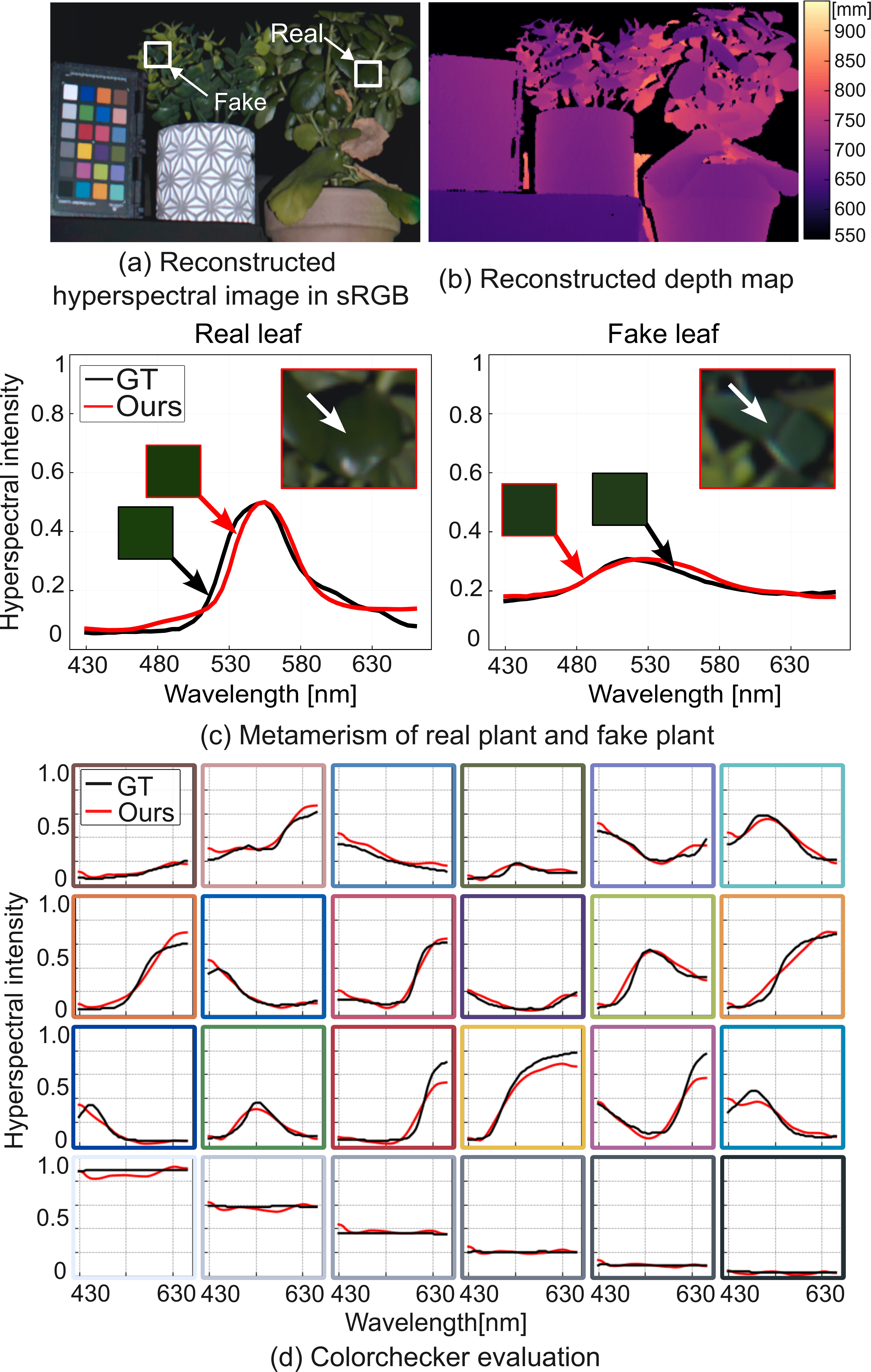}
\caption{\textbf{Evaluation on a ColorChecker and metameric samples.} (a) Reconstructed hyperspectral image in sRGB, (b) reconstructed depth map, (c) spectra of metameric samples (real and fake leaves), (d) reconstructed spectra of each color patch. }
\vspace{-5mm}
\label{fig:quantitative_reflectance}
\end{figure}

\paragraph{Hyperspectral 3D Reconstruction}
DSL enables accurate hyperspectral 3D imaging. We estimate a hyperspectral image with 46 channels, from 430\,nm to 660\,nm at 5\,nm intervals, along with a depth map. Figures~\ref{fig:teaser} and~\ref{fig:extra_scene} shows reconstruction results for two real-world scenes. 

\paragraph{Reconstruction of High-frequency Spectral Curves}
Figure~\ref{fig:peak_comparison} shows the results of our DSL in comparison with Li et al.~\cite{li2019pro}, state-of-the-art practical hyperspectral 3D imaging method.
We captured a scene containing nine bandpass spectral filters, each with a bandwidth of 10\,nm. DSL accurately identifies the center wavelengths and achieves an average FWHM of 18.8\,nm. 
In contrast, Li et al.~\cite{li2019pro} rely on broadband RGB illuminations of projector, resulting in a significantly broader FWHM of 83\,nm, due to the limited capability in differentiating high-frequency spectral features. This performance gap mainly originates from our use of first-order diffractions. We test the DSL without using the first-order term in Equation~\eqref{eq:hyperspectral_optimization2}. As expected, the resuting spectra is overly smooth with  50\,nm FWHM and cannot detect the high-peak spectral features. This is aligned with the results of Li et al.\cite{li2019pro}, demonstrating the importance of using first-order diffractions. 

\paragraph{Colorchecker and Metamerism}
Figure~\ref{fig:quantitative_reflectance} shows the spectral accuracy of our DSL measured on a ColorChecker and metameric fake and real leaves. 
For smooth spectral curves of color patches, DSL successfully reconstructs the hyperspectral intensity. 
Also, DSL enables telling clear difference between fake and real leaves. We obtained the ground-truth intensity using a spectroradiometer.

\begin{figure}[t]
\centering
\includegraphics[width=\columnwidth]{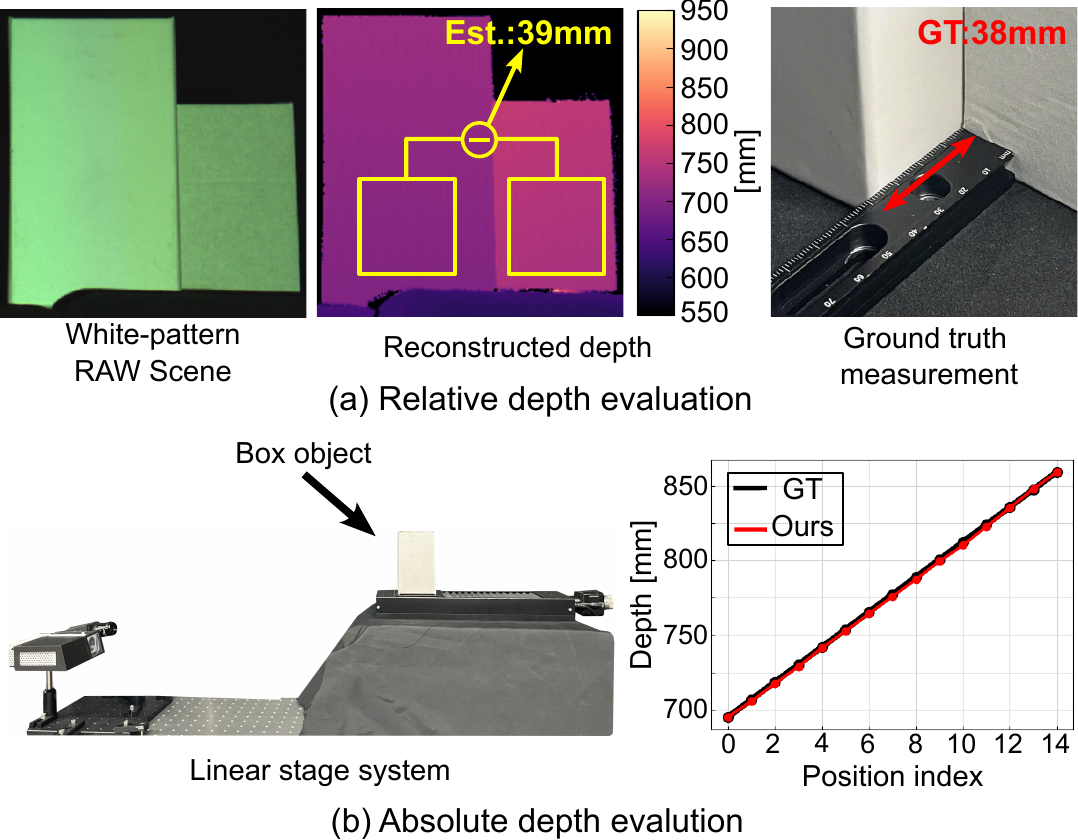}
\caption{\textbf{Depth accuracy.} {We estimate (a) the distance between two boxes and (b) the depth of the box mounted on a translation stage. We obtain the depth errors of 1\,mm for the relative-depth experiment and 1.35\,mm for the absolute depth experiment.} }
\vspace{-3mm}
\label{fig:quantitatiave_evaluation_depth}
\end{figure}

\paragraph{Depth Accuracy}
To assess the depth accuracy of our experimental prototype, we assess both relative and absolute depth errors. Figure~\ref{fig:quantitatiave_evaluation_depth}(a) show that DSL accurately estimates the distance between two boxes, with a marginal error of only 1\,mm. Figure~\ref{fig:quantitatiave_evaluation_depth}(b) evaluates the absolute depth error by capturing a planar object mounted on a linear translation stage (Thorlabs \#LTS150C). Across the working range of the translation stage with a 10\,mm step size, DSL achieves an average depth error of $1.35$\,mm.
Note that these experimental results are aligned with our synthetic experiments shown in Figure~\ref{fig:depth_reconstruction}.
\section{Conclusion}
In this paper, we introduced DSL, an accurate, low-cost, and compact hyperspectral 3D imaging method. Our dispersion-aware image formation, per-pixel hyperspectral 3D reconstruction, and calibration enables accurate hyperspectral 3D imaging.
DSL can be implemented in an affordable experimental prototype by using a diffraction grating adding sub-millimeter thickness at a cost of 10\,USD.
Our experimental prototype achieves depth error of 1\,mm and spectral FWHM of 18.8\,nm,  outperforming prior work on practical hyperspectral 3D imaging. We find that DSL makes a step towards practical hyperspectral 3D imaging for applications beyond computer vision and graphics. 

\paragraph{Limitations and Future Work}
While accurate and low-cost, our current experimental prototype takes around 10\,minutes to capture a scene, restricting it to static scenes.  
Also, the low intensity of first-order diffracted light limits the working depth range to be less than a meter.

For handling dynamic scenes and increasing depth range, we leave developing a light-efficient capture system and joint depth-spectrum reconstruction method as an interesting direction for future research.
Another unexplored direction is to find the optimal design of the diffraction pattern for efficient and accurate hyperspectral 3D imaging.
Recent differentiable optimization of imaging systems could be employed to achieve the goal. 
Lastly, the core principle of DSL can be applied to other spectral ranges beyond visible spectrum.

\paragraph{Acknowledgements}
Seung-Hwan Baek was supported by Korea NRF (RS-2023-00211658, 2022R1A6A1A03052954), Korea IITP MSIT (No.2019-0-01906, Artificial Intelligence Graduate School Program-POSTECH), and POSCO.
Felix Heide was supported by an NSF CAREER Award (2047359), a Packard Foundation Fellowship, a Sloan Research Fellowship, a Sony Young Faculty Award, a Project X Innovation Award, and an Amazon Science Research Award.

{\small
\bibliographystyle{ieeenat_fullname}
\bibliography{references}
}

\end{CJK}
\end{document}


\begin{CJK}{UTF8}{}
\CJKfamily{mj}

\title{
Dispersed Structured Light for Hyperspectral 3D Imaging\\
(Supplementary Information)}

\author{Suhyun Shin$^1$ ~ ~ ~
Seokjun Choi$^1$ ~ ~ ~
Felix Heide$^2$ ~ ~ ~
Seung-Hwan Baek$^1$ \\[2mm]
$^1$ POSTECH   ~ ~ ~   ~ ~ ~  $^2$ Princeton University\\
}

\maketitle

\hypersetup{linkcolor=black}
\noindent
In this supplemental document, we provide additional results and details of DSL in support of the findings from the main manuscript. 

\tableofcontents

\newpage
\section{Hyperspectral Acquisition Details}
\label{sec:implementation_details}
In this section, we outline the details of the capture settings utilized for hyperspectral reconstruction. To acquire data, we projected a sequence of 318 white scanline patterns. Each pattern is 5 pixels wide, with a 2-pixel shift. The entire data acquisition process was completed in approximately 9 minutes.

\subsection{HDR Imaging}
Our hyperspectral reconstruction method requires to attain both zero-order and first-order spectral information at valid intensity levels without saturation. This is enabled by using different capture settings which are merged into an HDR image. 

\begin{figure*}[ht]
	\centering
		\includegraphics[width=\linewidth]{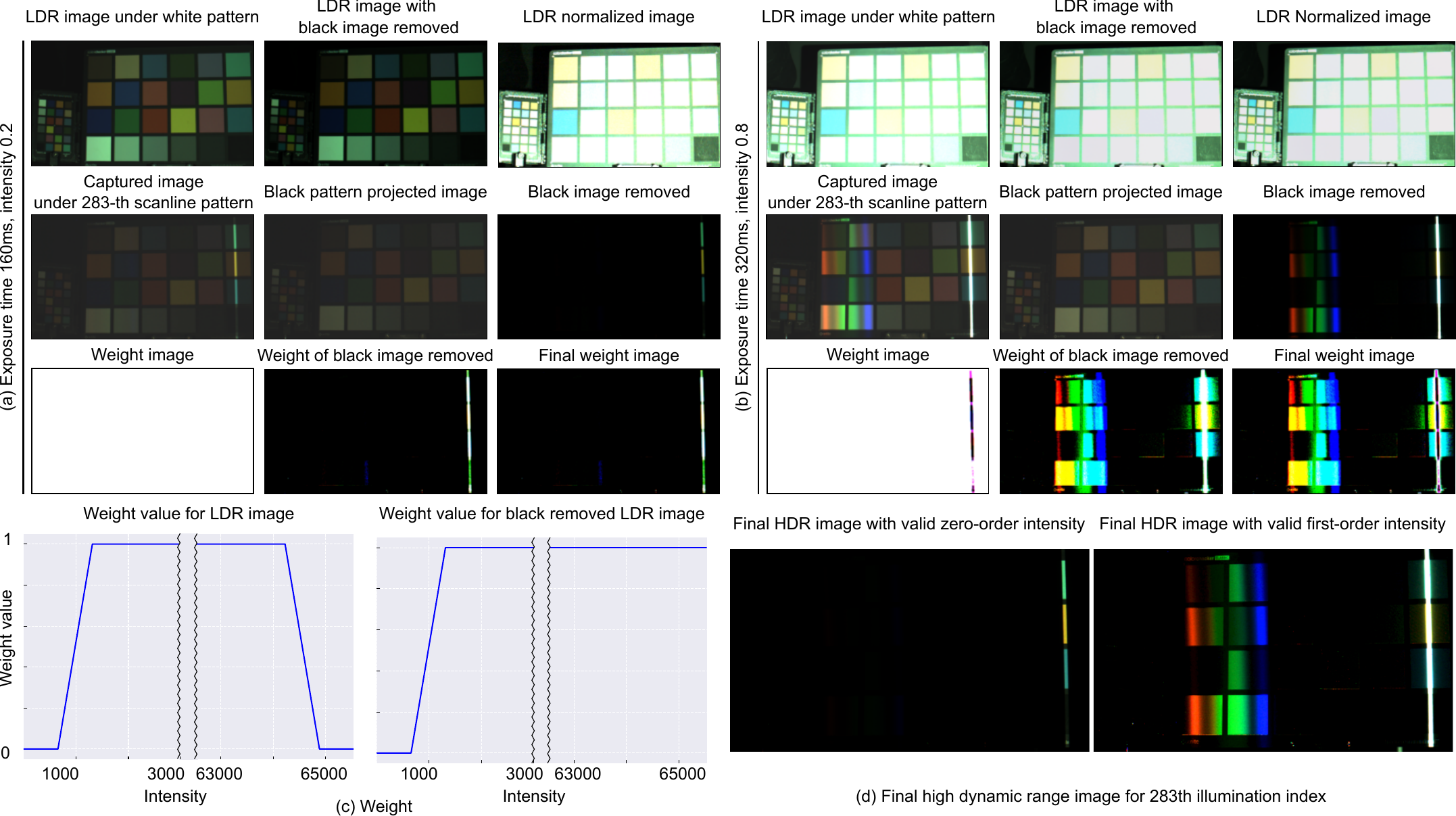}
  \caption{\textbf{HDR Imaging.} We show the procedure of HDR image for two different settings in (a) and (b). The first row of (a) and (b) shows the normalization process, next the black pattern removed LDR image and at last the weight images. (a) Exposure 160ms intensity 0.2, (b) exposure 320ms intensity 0.8, (c) Weight for original and black removed image, (d) final HDR image for $283^\text{th}$ illumination index.}
		\label{fig:hdr_imaging}
\end{figure*}

Due to the high-intensity difference between the zero-order and first-order light, capturing both orders with accurate intensity without saturation in a single capture setting is unattainable. Therefore, real scenes are captured under two settings: 160\,ms exposure time with 0.2 projector pattern intensity, 320\,ms time with 0.8 projector pattern intensity.
Once captured, we merge them into a HDR image. Figure~\ref{fig:hdr_imaging} show the HDR reconstruction process.

An HDR image is created by multiplying the total of weights with the total of radiance images generated by LDR images. Radiance images are black-level compensated. Intensity normalization is performed by normalizing the intensity of black patch Classic Color Checker in two different settings as Figure~\ref{fig:correspondence_model}(a) and (b) at the first row. We use different weights for the original LDR and black-removed LDR image to capture valid intensity information for both zero-order and first-order dispersed light as shown in Figure~\ref{fig:hdr_imaging}(c). The final weighted images are determined by finding the minimum between weights applied to the original LDR and the black-removed LDR image. This procedure is shown for two different settings in Figure~\ref{fig:hdr_imaging}(a) and (b), with the final HDR image for 283-th illumination index in Figure\ref{fig:hdr_imaging}(d).

\newpage

\begin{figure*}[ht]
    \centering
    \vfill
    \includegraphics[width=\textwidth]{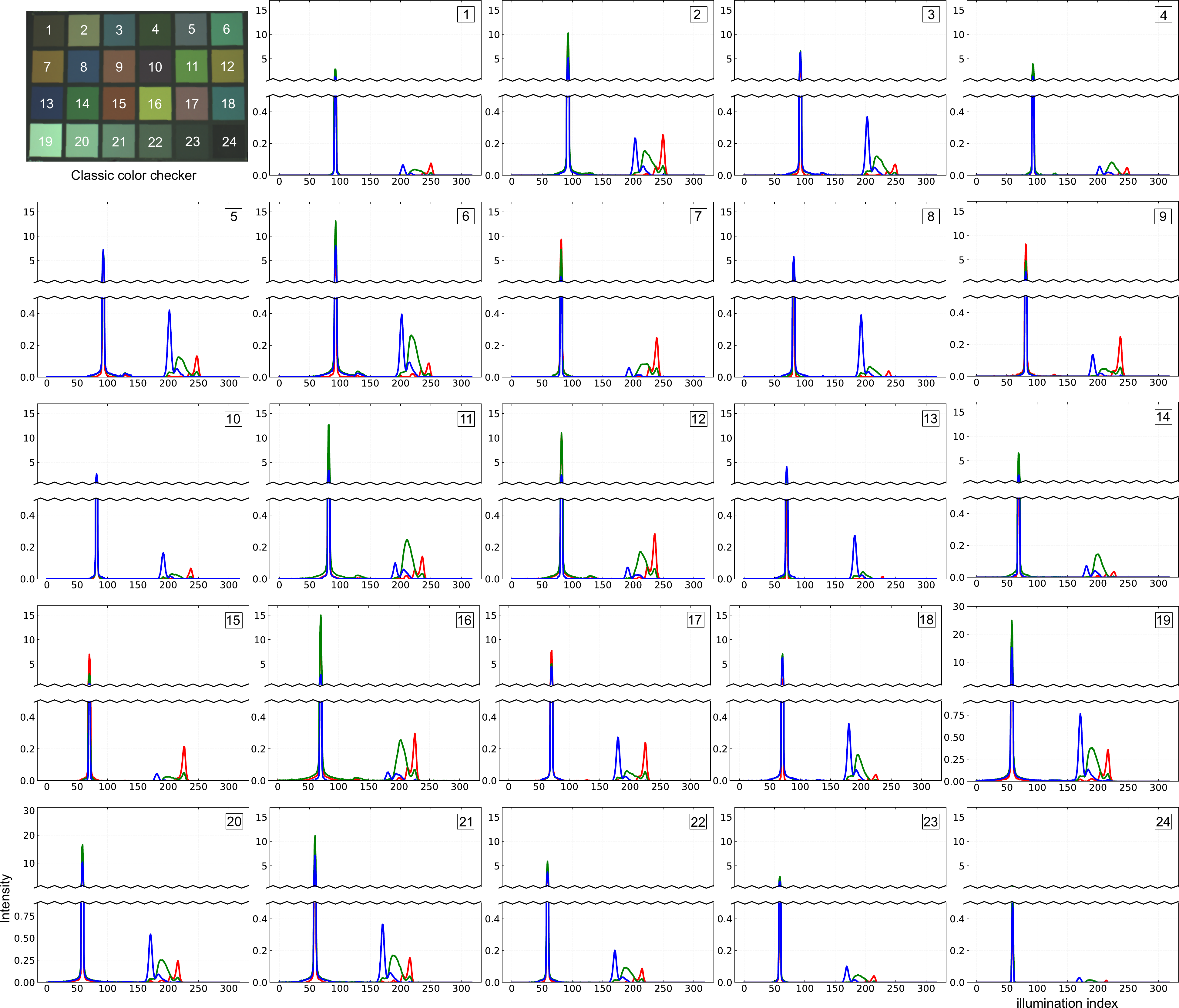} 
    \caption{Pixel intensity graph for central points of 24 Classic Color Checkers.}
    \label{fig:pdg}
    \vfill 
\end{figure*}

We finally generate pixel intensity graph which contains the valid RGB intensity value for both first-order dispersed light and zero-order light under scanline pattern. Figure \ref{fig:pdg} shows the pixel intensity graph of each 24 Color Checker center points. We assume that each pixel is illuminated by $430\text{nm}$ to $660\text{nm}$ wavelength light at $5\text{nm}$ interval with narrow bandwidth as the first-order dispersed light are shifted by a 2-pixel interval.

\newpage

\subsection{Handling Fast Modulation of Projector Light}
Our method requires accurate capture of the true RGB color and intensity of both zero-order and first-order dispersed light. A key issue to address in our method is the illumination pulsing of red, green, and blue LEDs by the projector. This pulsing happens fast enough making our eyes naturally see a mix of these colors as a single full-color image. However, if the camera exposure setting is too low, it measure these colors separately instead of blended together, which could lead to inaccurate capture. While synchronization can resolve this issue, in our prototype, we set the minimum exposure time to 160\,ms to avoid any incorrect color captures and make sure our imaging results are reliable and accurate. 

\section{Part List of the Experimental Prototype}
We list all parts used to build the experimental prototype
system in Table~1.

\begin{table*}[!htb]
\centering
\label{tab:table1}
\begin{tabular}{clcc}
\hline
\textbf{Item \#} & \textbf{Part description} & \textbf{Quantity} & \textbf{Model name} \\
\hline
1 & RGB Camera & 1 & FLIR GS3-U3-32S4C-C \\
2 & RGB Projector & 1 & LG PH30N \\
3 & Objective lens & 1 & Edmund \#33-303\\
4 & Diffraction grating sheet & 1 & Edmund 158 \#54-509 \\
5 & Holder & 1 & 3D printed  \\
\hline
\end{tabular}
\caption{Part list of our imaging system.}
\end{table*}

\section{Calibration Details}
We perform geometric calibration of the camera and the projector without attaching the diffraction grating~\cite{taubin20143d}. In the following, we describe the details on radiometric calibration.

\begin{figure*}[ht]
    \centering
    \vfill
    \includegraphics[height=0.8\textheight]{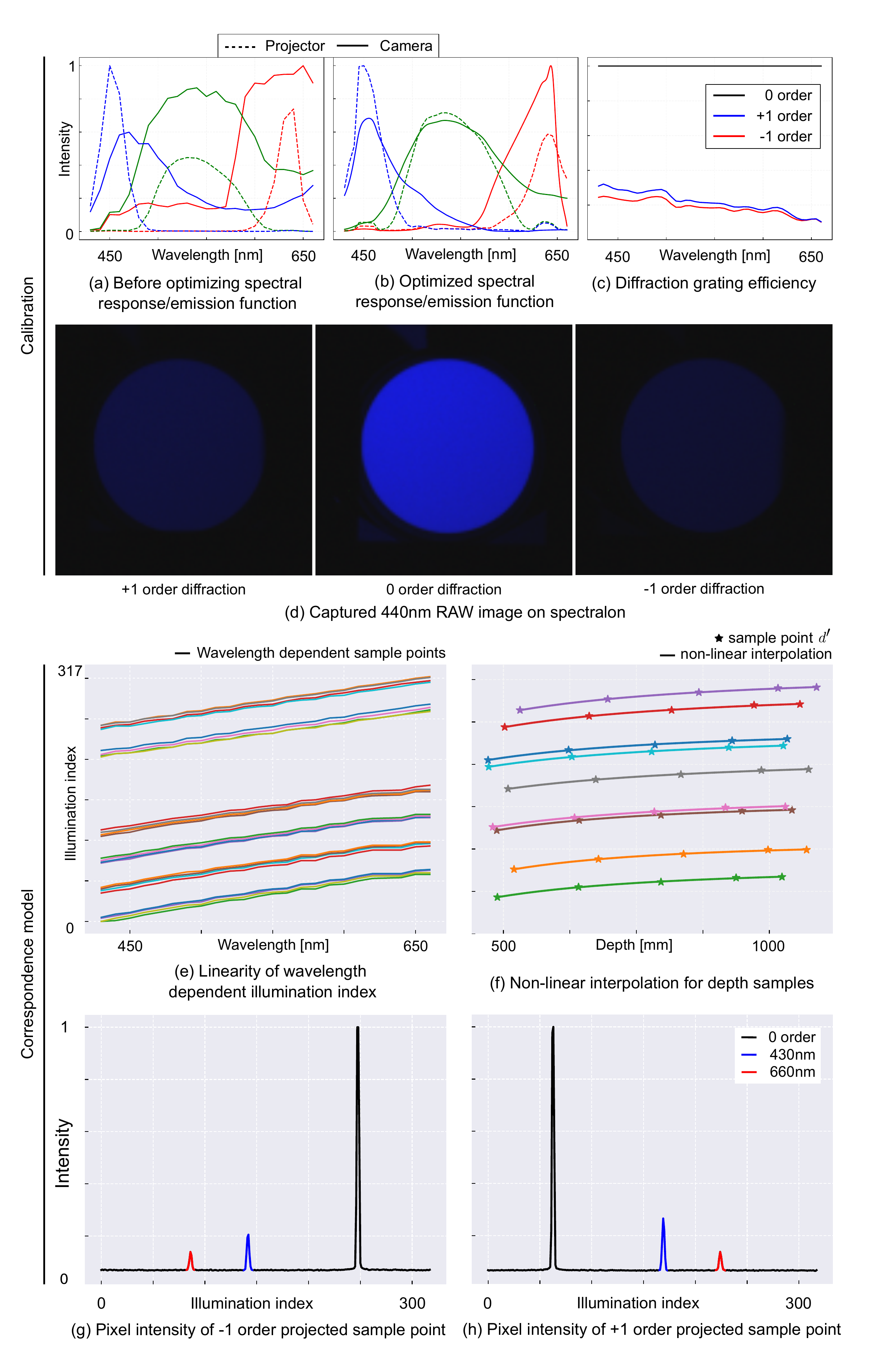} 
    \caption{\textbf{Calibration and Correspondence Model.} (a) The spectral response/emission function before optimization, (b) spectral response/emission function after optimization, (c) diffraction grating efficiency, (d) captured 440nm raw image on a spectralon for each $m$ order, (e) linearity of wavelength dependent illumination index, (f) non-linear interpolation for each depth samples. (g) and (h) shows the pixel intensity graph for 0 order and 430\,nm 660\,nm first orders. (g) is the -1 order projected sample point, and (h) is the +1 order projected sample point pixel intensity.}
    \label{fig:correspondence_model}
    \vfill 
\end{figure*}

\subsection{Refinement of Spectral/Emission Function}
We obtain projector emission function by projecting red, green, and blue dots onto a Spectralon target, measuring the reflected radiance with a spectroradiometer (JETI Specbos 1211), and normalizing the results with Spectralon reflectance. For the camera response function, we use the data provided by the camera manufacturer. We show the initial parameters of camera spectral and projector emission function in Figure~\ref{fig:correspondence_model}(a). Next, we refine the spectral response and emission functions within the integrated DSL system. This refinement aims to minimize the discrepancy between actual intensity $\textbf{I}_m$ and simulated intensity under scanline patterns as Equation~\eqref{eq:optimization_resp}. We utilized the simulated pixel intensity graphs of 21 central points $p \in \mathcal{P}$ of the Classic Color Checker, employing the given ground truth hyperspectral reflectance $H(p,\lambda)$.

\begin{align}
        \underset{\mathbf{\Omega's}}{\text{argmin}} \sum_{p \in \mathcal{P}} \| \textbf{I}_m -&
         \Omega^{\text{cam}}_{c, \lambda}  H(p,\lambda)\eta_{m,\lambda}L(q_{m,\lambda}, \lambda)\|_2^2 \nonumber \\ 
         +& w(\| \nabla_\lambda \Omega^{\text{cam}}_{c, \lambda}\|_2^2 + \| \nabla_\lambda \Omega^{\text{proj}}_{c', \lambda}\|_2^2),
    \label{eq:optimization_resp}
\end{align}
where $\nabla_\lambda$ is the gradient operator along the spectral axis, and we set the spectral weight $w$ as 0.005. The result of optimized spectral response and emission function is shown in Figure~\ref{fig:correspondence_model}(b). 

\subsection{Diffraction Grating Efficiency}
We calibrate the diffraction grating efficiency $\eta_{m, \lambda}$ by measuring the intensity of each $m$-order diffracted light. We place spectral band pass filters at 10\,nm intervals from 430\,nm to 660\,nm to capture at each wavelength $\lambda$ projected onto a Spectralon sample. We show a captured Spectralon sample at wavelength 440\,nm for first order and zero-order diffracted lights in Figure~\ref{fig:correspondence_model}(d).
The diffraction efficiency is then computed as the intensity ratio of each first-order wavelength measurement over the zero-order intensity. The calibrated result of diffraction efficiency is shown in Figure~\ref{fig:correspondence_model}(c). 

\section{First-order Correspondence Model}
{We generated a first-order correspondence model with better time efficient compared to iterative analytic correspondence estimation. Such iterative methods require 3\, minute optimization process per each scene, whereas ours take 2.5\, seconds. In addition, it suffers from inaccuracy due to deformability and pose ambiguity of diffraction grating film.}

\subsection{Data Acquisition} We describe the details of data-driven correspondence model for first-order diffractions ($m$ = -1 or 1). The ultimate goal of our correspondence model is to get the exact indices of the patterns that illuminate pixel $p$. To achieve this, we acquired images of flat surfaces at five different depth positions, ranging from $50$cm to $100$cm each under scanline illumination patterns. These images were each captured using seven spectral bandpass filters 430\,nm, 600\,nm, 610\,nm, 620\,nm, 640\,nm, 650\,nm, 660\,nm. Figure~\ref{fig:correspondence_model}(g) and (h) depicts an example of pixel intensity graph of -1 order and +1 order projected pixel. Here, we display three peaks, each corresponding to zero-order light and first order lights with wavelength 430\,nm and 660\,nm. For a subset of grid sampled camera pixels $p' \in \mathbb{R}^{48\text{x}85}$, we identified the corresponding projector pixel $q'_{m,\lambda}$ from the captured images. Figure~\ref{fig:correspondence_model}(e) and (f) illustrates that interpolation methods for wavelength, camera pixel $p'$ and depth $z'$ can be utilized to address samples not included in sets $\mathbb{S}_\lambda$, $\mathbb{S}_p$, $\mathbb{S}_z$. 

\subsection{Interpolation}
For each camera pixel sample $p'$ and wavelength $\lambda'$, we model the corresponding projector pixel $q'_{m,\lambda}$ relative to depth $z$ by fitting the samples to 
\begin{equation}
\label{eq:power_function}
q_{m,\lambda} = \alpha z^ {\beta} + \gamma,
\end{equation}
where $\alpha$, $\beta$, and $\gamma$ are the parameters, estimated using curve fitting in MATLAB non-linear least squared method with the power of 2. Figure~\ref{fig:correspondence_model}(f) shows the interpolated results of 5 different depth sample point $z'$ for random pixel $p'$ and wavelength $\lambda'$. 
We experimentally show that we can use linear interpolation for wavelength and cubic interpolation for projector pixel $q$ to handle samples which are not included in sets $\mathbb{S}_\lambda$ and $\mathbb{S}_p$ in Figure~\ref{fig:correspondence_model}(e). With this procedure we define our correspondence function for first-order diffractions.

\subsection{Valid Orders} 
Not every region of a scene is illuminated by both plus and negative first-order light due to the limited degree of freedom of the patterns. The valid first-order projector pattern index can be obtained as
\begin{equation}
\delta_{1st,i} = \texttt{is\_valid}\left(\psi\left(p,z,\{-1,1\},\lambda_i\right)\right),
    \label{eq:correspondence_first}
\end{equation}
We define the valid first orders by comparing the illumination index of zero-order and first-order shown in Figure~\ref{fig:correspondence_model}(g) and (h). The highest illumination intensity corresponds to the zero-order, while the subsequent intensity level represents the first-orders. Here, we only plot the peak of two wavelengths, 430\,nm and 660\,nm. It is apparent that when zero-order pattern index exceeds the first order, such pixel $p$ is illuminated by -1 order light as Figure~\ref{fig:correspondence_model}(g) and vice versa +1 order as Figure~\ref{fig:correspondence_model}(h). Therefore, by utilizing Equation \eqref{eq:correspondence_first} we determine the $m$ order to be applied for each sample point.

\section{Validation of Binary Decoding}
{DSL’s depth accuracy is bounded by the sub-millimeter accuracy of conventional SL~\cite{gupta2011structured}. We experimentally achieve 1 mm depth error and demonstrate the validity of conventional SL decoding for DSL.}
We start by rewriting Equation~\eqref{eq:intensity_min_max} in the main paper as follows:

\begin{align}
        I^\text{on}_\text{min}(p,c)&=\sum_{\lambda } \Omega^{\text{cam}}_{c, \lambda} H(p,\lambda) \eta_{0,\lambda} \sum_{c'} \Omega^{\text{proj}}_{c', \lambda},\\
        I^\text{off}_\text{max}(p,c) &= \sum_{\lambda } \Omega^{\text{cam}}_{c, \lambda} H(p,\lambda) \sum_{m={1,-1}} \eta_{m,\lambda} \sum_{c'} \Omega^{\text{proj}}_{c', \lambda}.
    \label{eq:intensity_min_max}
\end{align}
We can acquire each intensity $I^\text{on}_\text{min}(p,c)$ and $I^\text{off}_\text{max}(p,c)$ by defining the maximum pixel intensity which is projected by all m orders and wavelength $\lambda$ under structured light as

\begin{equation}
        I(p,c) = \sum_{\lambda \in \Lambda} \Omega^{\text{cam}}_{c, \lambda}  H(p,\lambda)  \sum_{m=-1} \eta_{m,\lambda} L(q_{m,\lambda}, \lambda) 
    \label{eq:pixel_intensity}
\end{equation}
It is clear that the minimum captured intensity with only zero-order structured light $I^\text{on}_\text{min}(p,c)$ can be derived by Equation~\eqref{eq:intensity_min} by only leaving the order m = 0 valid.
\begin{align}
        I(p,c) &= \sum_{\lambda \in \Lambda} \Omega^{\text{cam}}_{c, \lambda}  H(p,\lambda)  \sum_{m=-1} \eta_{m,\lambda} L(q_{m,\lambda}, \lambda)  \nonumber \\
        &\geq \sum_{\lambda \in \Lambda} \Omega^{\text{cam}}_{c, \lambda}  H(p,\lambda)  \eta_{0,\lambda} L(q_{0,\lambda}, \lambda) \nonumber \\
        &=\sum_{\lambda \in \Lambda} \Omega^{\text{cam}}_{c, \lambda}  H(p,\lambda)  \eta_{0,\lambda} \sum_{c'} \Omega^{\text{proj}}_{c', \lambda} P(q_{0,\lambda},c') \nonumber\\
        &=\sum_{\lambda \in \Lambda} \Omega^{\text{cam}}_{c, \lambda}  H(p,\lambda)  \eta_{0,\lambda} \sum_{c'} \Omega^{\text{proj}}_{c', \lambda}= I^\text{on}_\text{min}(p,c) \nonumber\\
    \label{eq:intensity_min}
\end{align}
The maximum captured intensity $I^\text{off}_\text{max}(p,c)$ is when zero-order structured light does not illuminate the pixel and the first-order does. This can be written as Equation \eqref{eq:intensity_max} by only adding the first orders m = -1 and 1.
\begin{align}
I^\text{off}_\text{max}(p,c) &=  \sum_{\lambda \in \Lambda} \Omega^{\text{cam}}_{c, \lambda}  H(p,\lambda)\sum_{m=1,-1} \eta_{m,\lambda} L(q_{m,\lambda}, \lambda) \nonumber \nonumber\\
&= \sum_{\lambda \in \Lambda} \Omega^{\text{cam}}_{c, \lambda}  H(p,\lambda)  \sum_{m=\{1,-1\}} \eta_{m,\lambda} \sum_{c'} \Omega^{\text{proj}}_{c', \lambda} P(q_{m,\lambda},c') \nonumber \\
&= \sum_{\lambda \in \Lambda} \Omega^{\text{cam}}_{c, \lambda}  H(p,\lambda) \sum_{m=\{1,-1\}} \eta_{m,\lambda} \sum_{c'} \Omega^{\text{proj}}_{c', \lambda} 
\label{eq:intensity_max}
\end{align}

Since $\eta_{0, \lambda}\sum_{c'} \Omega^{\text{proj}}_{c', \lambda} > \sum_{m=\{1,-1\}} \eta_{m,\lambda}\sum_{c'} \Omega^{\text{proj}}_{c', \lambda}$ holds for every wavelength $\lambda$ shown in Figure ~\ref{fig:correspondence_model} (c), we can validate conventional binary-code decoding for DSL.

\section{Additional Discussion}
We presented an accurate hyperspectral 3D imaging technique achieved through DSL. This method significantly surpasses previous methods that utilized RGB cameras and projectors. The difference in our approach is the use of a diffraction grating sheet, which is both practical and cost-effective. Our experimental prototype reconstructs depth with a minimal error of 1\,mm. Additionally, it acheives accurate hyperspectral imaging with a spectral FWHM of 18.8\,nm. This level of accuracy and efficiency marks a notable improvement in the field of hyperspectral 3D imaging.

\subsection{Efficiency and Optimized Patterns} Although our DSL capture system is accurate while low-cost and compact, there exist limitations in terms of efficiency. Currently, our reconstruction method relies on an optimization method for each single scene. This necessitates a separate optimization for every distinct scene, adding complexity. Additionally, for the reconstruction of hyperspectral 3D imaging, our current method projects 40 binary-code patterns along with 318 white scanline patterns. While this technique facilitates accurate reconstruction results, it also comes with extended processing times. To address this issue, integrating differentiable optimization into the imaging systems may be employed. This would enable to reduce the total pattern numbers by defining an optimal pattern for DSL. Also, enhancing our approach by avoiding the need to optimize the hyperspectral image for each scene significantly increases efficiency.

\subsection{Global Illumination}
{We leave designing optimal DSL patterns robust to global illumination as future work. This will overcome the fundamental problem of structured light caused by subsurface scattering, defocus, and interreflection which lowers depth accuracy.}

\subsection{Imaging System}
Our prototype requires a complex and time-consuming data-driven calibration method. We capture each depth levels under 318 scanline patterns and 40 binary-codes patterns. While the proposed method relies on a thin and flexible diffractive film, future imaging systems could incorporate a fixed non-flexible diffraction grating. Such a design would be less prone to errors in simulating diffracted rays, and allows for a simplified calibration process improving overall system performance.

\begin{figure}[t]
    \begin{minipage}{\linewidth}
        \centering
        \includegraphics[width=0.63\textwidth]{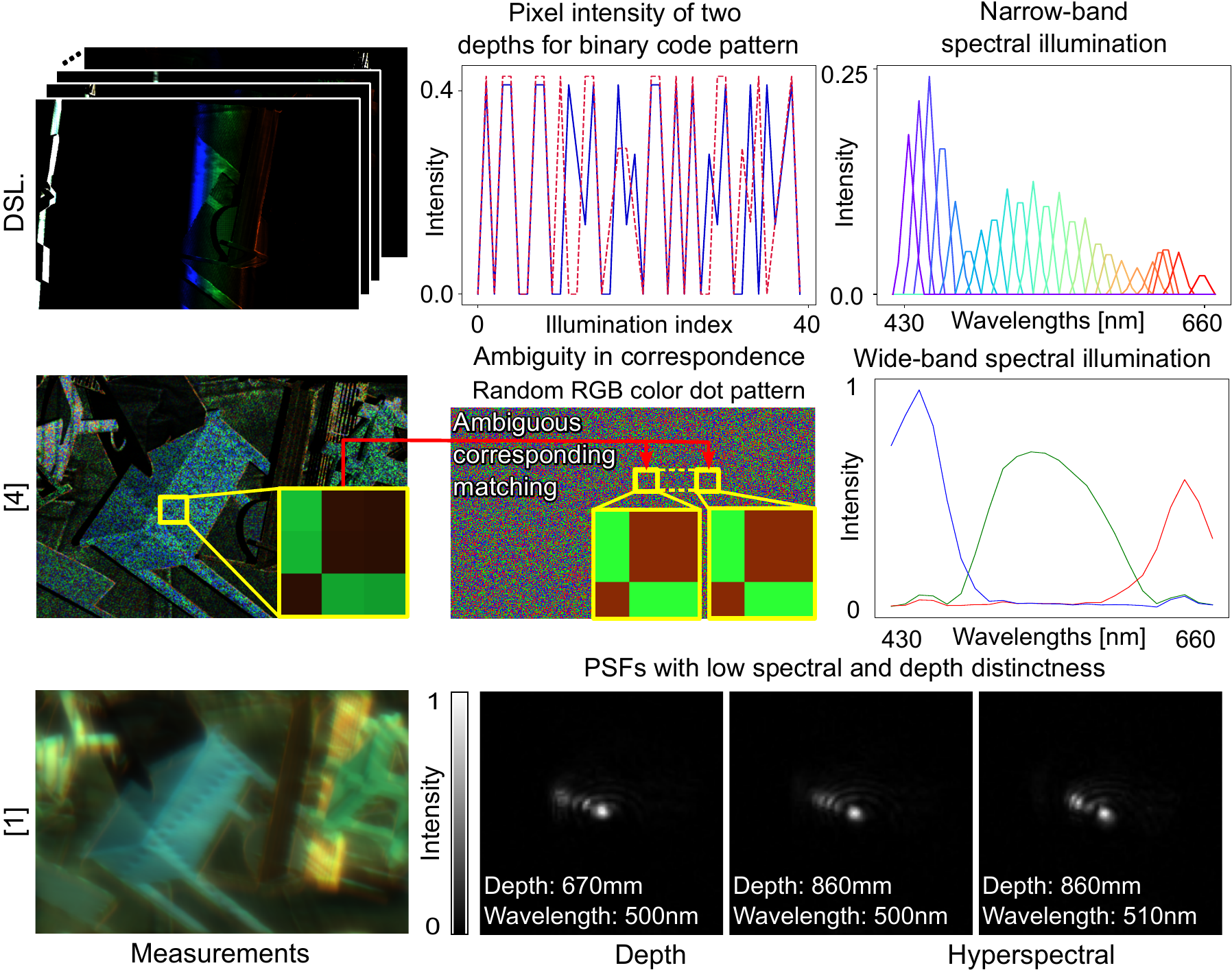}
    \end{minipage}

    \vspace{4mm}
    
    \begin{minipage}{\linewidth}
        \centering
        \resizebox{0.6\linewidth}{!}{%
            \tiny 
             \begin{tabular}{c|c|c}
        & Depth accuracy & Spectral accuracy \\ \hline
        DSL & \cellcolor[HTML]{D6F6D5} \begin{tabular}[c]{@{}c@{}}Per-pixel analytic \\ reconstruction using \\ structured-light decoding\end{tabular} & \cellcolor[HTML]{D6F6D5} \begin{tabular}[c]{@{}c@{}}Per-pixel analytic\\ reconstruction using \\ narrow-band hyperspectral illum.\end{tabular} \\
        \hline
        [4] & \cellcolor[HTML]{FDEFB2} \begin{tabular}[c]{@{}c@{}}Global learning-based \\ reconstruction using\\ stereo correspondence matching\end{tabular} & \cellcolor[HTML]{FFE6EE} \begin{tabular}[c]{@{}c@{}}Global learning-based \\ reconstruction using \\ wide-band RGB illum.\end{tabular} \\
        \hline
        [1] & \multicolumn{2}{c}{\cellcolor[HTML]{FFE6EE} \begin{tabular}[c]{@{}c@{}}Global learning-based \\ reconstruction using deconvolution \\ with low-frequency spectral-depth PSFs\end{tabular}} \\
        \end{tabular}%
        }
    \end{minipage}
    
    \caption{\textbf{DSL vs. [4, 23].} As shown in the bottom table and visualization, DSL achieves high spectral and depth accuracy by utilizing per-pixel narrow-band hyperspectral illumination and structured light decoding. The other methods~[4,23] suffer from wide-band illumination and low-frequency blur cues.}
    \label{fig:DSLvsSOTA}
\end{figure}

\begin{figure}
    \vspace{4mm}
    \centering
    \includegraphics[width=0.6\textwidth]{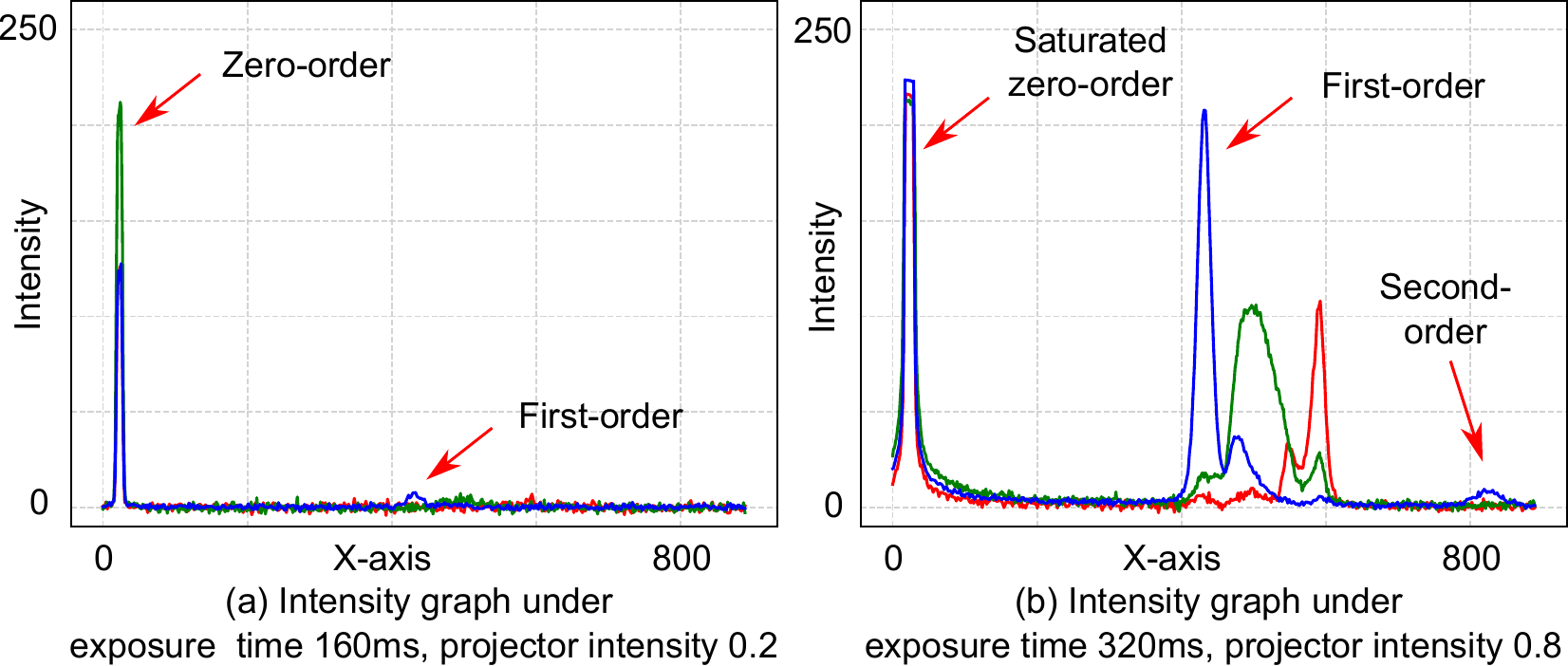}
    \caption{Intensity graph comparing 0/1/2 order diffraction. (a) First-order diffraction has 4\% intensity compared to zero-order, and (b) second-order has 4\% intensity compared to first-order. }
    \label{fig:Second-order diffraction}
\end{figure}

\newpage
\subsection{Comparison with Practical Hyperspectral 3D Imaging}
{We compared DSL with \cite{li2019pro} in our main paper by re-implementing it, thanks to its non-learning-based reconstruction. However, re-implementing and training the other two learning-based methods \cite{li2022deep}, \cite{baek2021single} are challenging due to their missing network-reconstruction code. Instead in Figure~\ref{fig:DSLvsSOTA}, we show that only DSL provides distinct cues while others result in less distinctness measurements.}

\subsection{Effect of Second-order Diffraction}
{We experimentally show that we can neglect second-order diffraction in case they are within our FoV. Figure~\ref{fig:Second-order diffraction} shows that intensity of second-order diffracted light has only 4\% of first-order intensity, which is too small to affect reconstruction results within our dynamic range. For this experiment, we require the second-order diffraction to be captured within FoV.}

\section{Hyperspectral Reconstruction Details}

In this section, we provide details of our hyperspectral optimization. We aim to optimize \ a highly accurate hyperspectral image $\textbf{H}$, with our proposed optimization method for 1000 epochs. This entire optimization process is completed within 3 minutes on average.

\subsection{Optimization Time and Weight Values}
\begin{figure*}[ht]
	\centering
		\includegraphics[width=\linewidth]{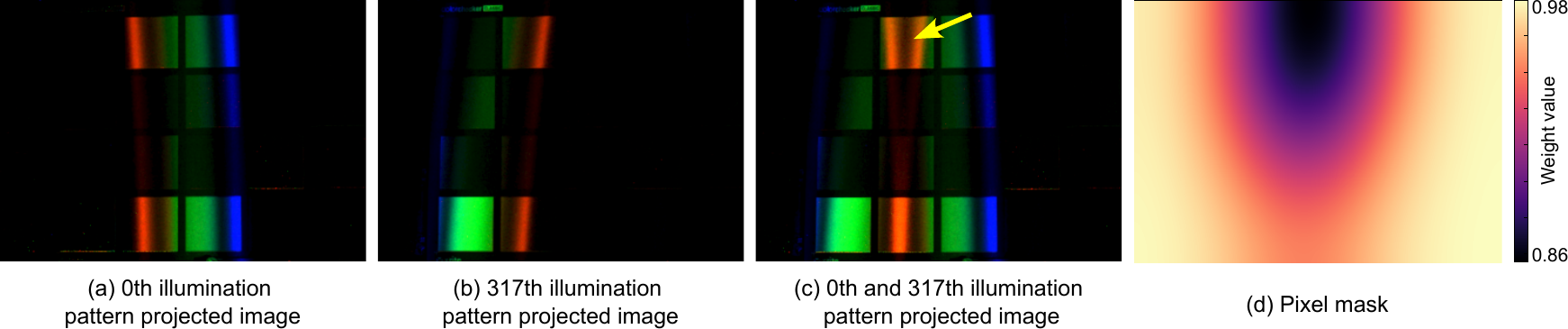}
  \caption{\textbf{Details on Optimization.} We capture a Color Checker scene under scanline pattern and visualize the pixel mask for first-order light unreached regions. (a) 0-th illumination pattern projected image, (b) 317-th illumination pattern projected image, (c) 0-th and 317-th illumination pattern projected image, (d) pixel mask.}
		\label{fig:optimization}
\end{figure*}

Some pixels have incomplete $\textbf{I}_{m}$ because they do not receive long-wavelength first-order light at certain depths. Figure \ref{fig:optimization}(a)-(c) shows HDR image of 0-th and 317-th illumination projected on to the scene. Starting from \ref{fig:optimization}(a), this shows the 0-th illumination scanline pattern projected to the scene and \ref{fig:optimization}(b) projected by 317-th pattern. Figure \ref{fig:optimization}(c) shows the gap where long-wavelength first-order light does not reach at a certain region. As sure, we yield a pixel mask with incomplete intensity and apply Gaussian blur to it to obtain the weight of spectral prior $\kappa_m$, where $\kappa_1$ is the first-order weight and zero-order weight $\kappa_0$ is $1 - \kappa_1$. Figure~\ref{fig:optimization} (b) shows the spatially-varying weight map of $\kappa_1$ for the specific Color Checker scene.

\section{Details on Depth Evaluation}
We discuss details of the depth evaluation and the reconstruction results of real and synthetic binary-code projected scene.

\subsection{Synthetic Dataset}
For depth estimation we used binary-code structured light patterns and assess the accuracy in different noise level in DSL and conventional SL method. We experiment this by using our synthetic dataset which simulates both first and zero-order lights.

\subsubsection{Correspondence for Zero-order Diffraction}
For zero-order diffracted light ($m=0$), the camera-projector correspondence can be earned as if diffraction grating in front of the projector is not present~\cite{Hecht:2001:Optics}. Therefore, it is the same as that of conventional structured light.
\begin{align}
q &= \psi\left(p,z,0,\lambda\right)=\texttt{proj}\left(\texttt{unproj}\left(p, z\right),0,\lambda \right),\\ 
{\mathbf{S}} &= \texttt{unproj}(p, z) = z(\mathbf{E}^{\text{cam}})^{-1}(\mathbf{K}^{\text{cam}})^{-1} \texttt{homo}(p), \label{eq:unproj}\\
q &= \texttt{proj}({\mathbf{S}},0,\lambda) = \texttt{homo}^{-1}(\mathbf{K}^{\text{proj}} \mathbf{E}^{\text{proj}} {\mathbf{S}}), \label{eq:proj}
\end{align}
where {$\mathbf{S}$} is the unprojected 3D point, $\mathbf{E}^{\text{cam}}$ and $\mathbf{K}^{\text{cam}}$ are the extrinsic and intrinsic matrices of the camera, respectively, and $\mathbf{E}^{\text{proj}}$ and $\mathbf{K}^{\text{proj}}$ are the extrinsic and intrinsic matrices of the projector. 
Image distortion is considered, however, omitted in Equations for simplicity.
The function $\texttt{homo}()$ converts ordinary 2D pixel coordinate to  homogeneous coordinate, and $\texttt{homo}^{-1}()$ performs its inverse conversion.

\subsubsection{Correspondence for First-order Diffraction}
\begin{figure*}[ht]
    \centering
    \vfill
    \includegraphics[width=0.93\textwidth]{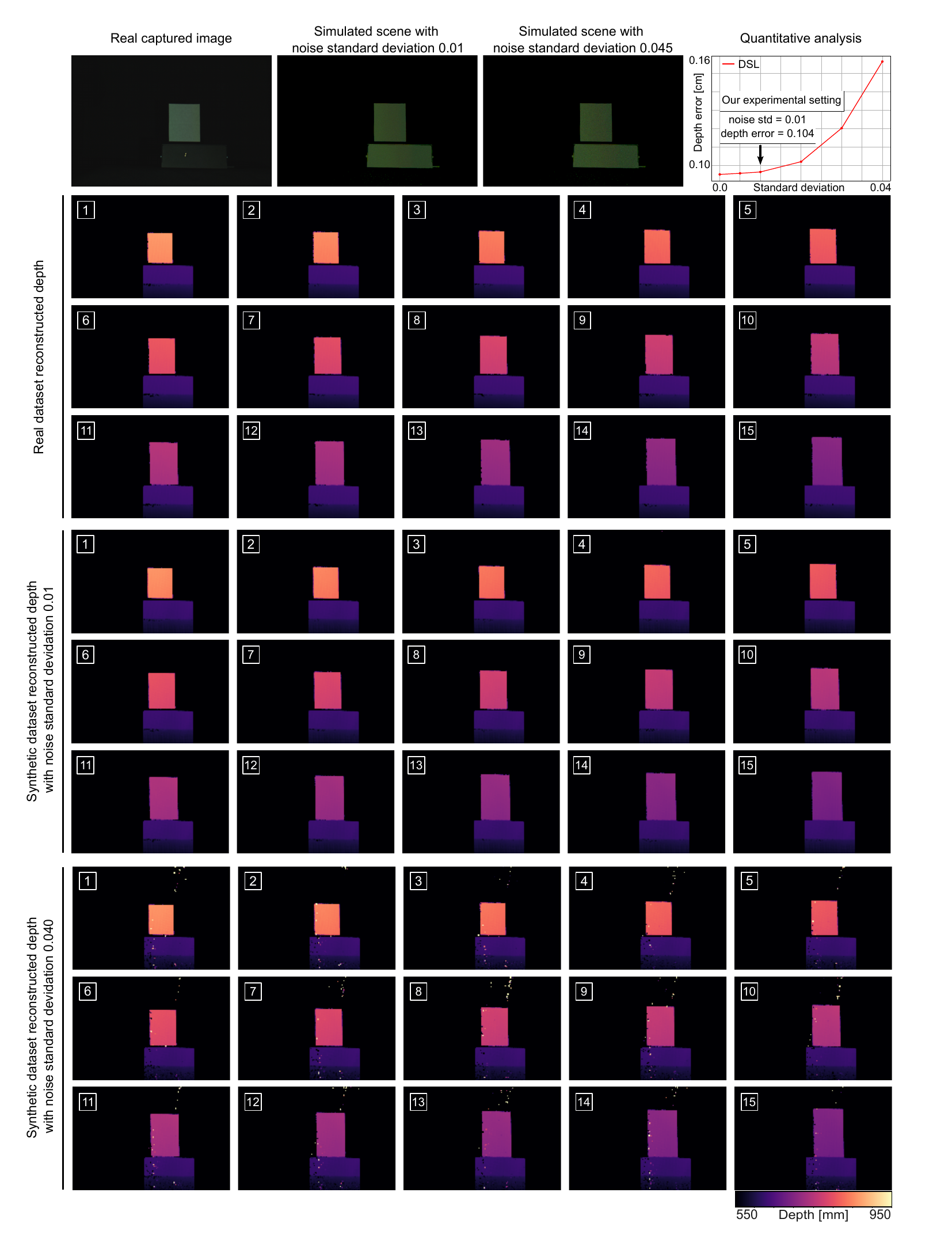} 
    \caption{\textbf{Additional Depth Evaluation.} The first row shows from left to right, captured images of an object on a linear translation stage in a real scene, a synthetic scene with noise level 0.01, and 0.04. Next we display the reconstructed depth for all 15 position index.}
    \label{fig:depth_evaluation}
    \vfill 
\end{figure*}

Different from zero-order diffraction, we obtain the correspondence between camera pixel $p$ and m order $\lambda$ wavelength projector pixel $q_{m, \lambda}$ with our data-driven correspondence model as Equation \eqref{eq:correspondence_model}.

\begin{equation}
    q_{m, \lambda} = \psi\left(p,z,m,\lambda\right)
    \label{eq:correspondence_model}
\end{equation}
We show the simulated box object in Figure~\ref{fig:depth_evaluation} the under white pattern.

\subsection{Additional Details on Depth Evaluation}
To evaluate the depth accuracy of our experimental prototype, we capture a planar object mounted on a linear translation stage(Thorlabs \#LTS150C) shown in Figure 9(b) of our main paper. Additionally, we simulate this in various Gaussian noise standard deviation [0.0, 0.005, 0.01, 0.02, 0.03, 0.04] to assess reconstruction quality in simulation. 
We show the reconstructed depth result of real captured binary-code images and synthetic datasets with different noise levels. In Figure~\ref{fig:depth_evaluation}, we show images with 0.01 and 0.04 noise level where 0.01 corresponds to representative noise in our hardware measurements.

\section{Additional Details on Hyperspectral Evaluation}

\subsection{Hyperspectral Reconstruction}
\begin{figure*}[ht]
    \centering
    \vfill
    \includegraphics[width=\textwidth]{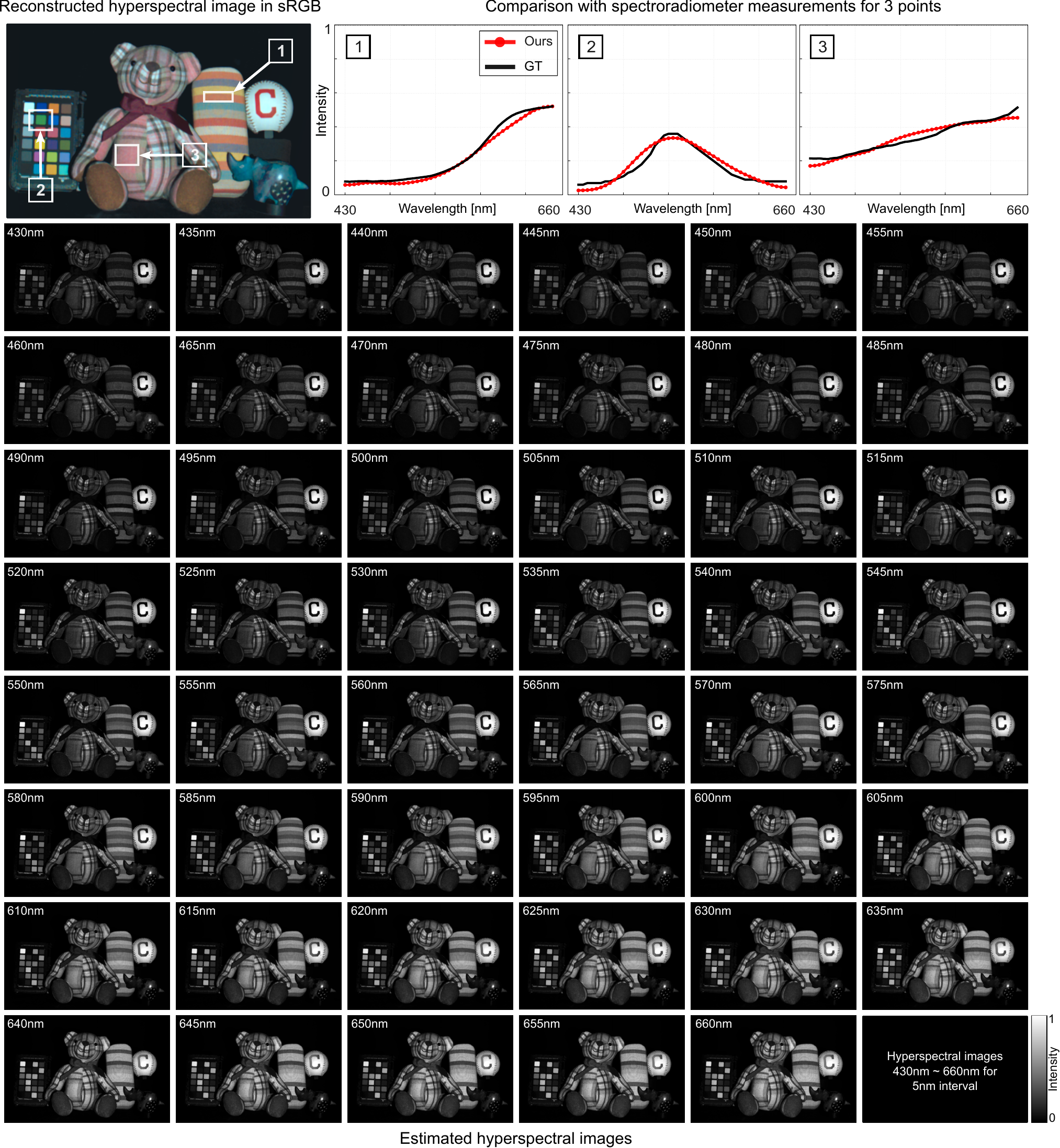} 
    \caption{\textbf{Hyperspectral Imaging.} We show the reconstructed hyperspectral image in sRGB, the comparison with spectrometer measurements for 3 points and detailed hyperspectral image for 5 nm interval.}
    \label{fig:hyperspectral_scene1}
    \vfill 
\end{figure*}

\begin{figure*}[ht]
    \centering
    \vfill
    \includegraphics[width=\textwidth]{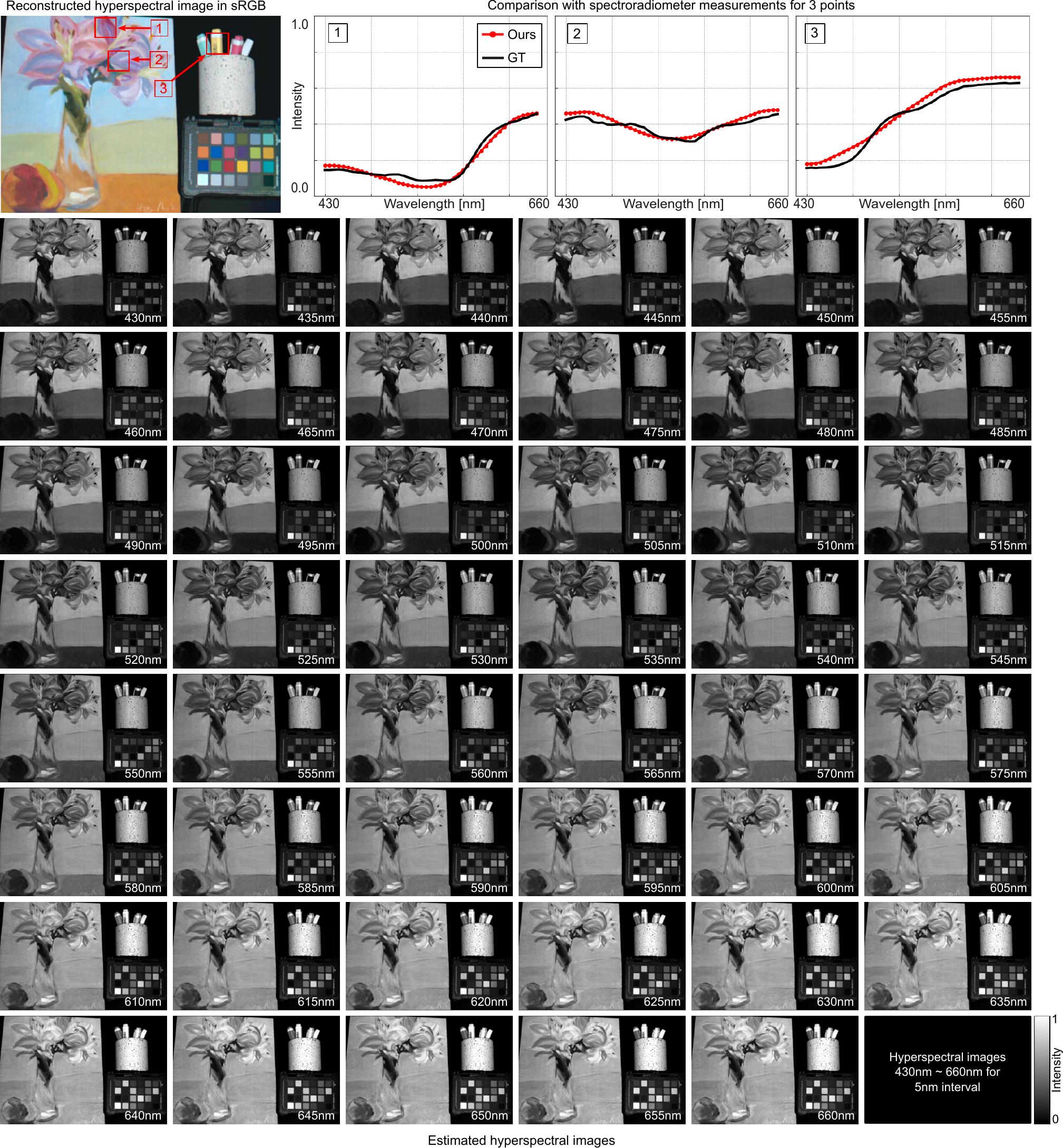} 
    \caption{\textbf{Hyperspectral Imaging.} We show the reconstructed hyperspectral image in sRGB, the comparison with spectrometer measurements for 3 points and detailed hyperspectral image for 5 nm interval.}
    \label{fig:hyperspectral_scene2}
    \vfill 
\end{figure*}

\begin{figure*}[ht]
    \centering
    \vfill
    \includegraphics[width=\textwidth]{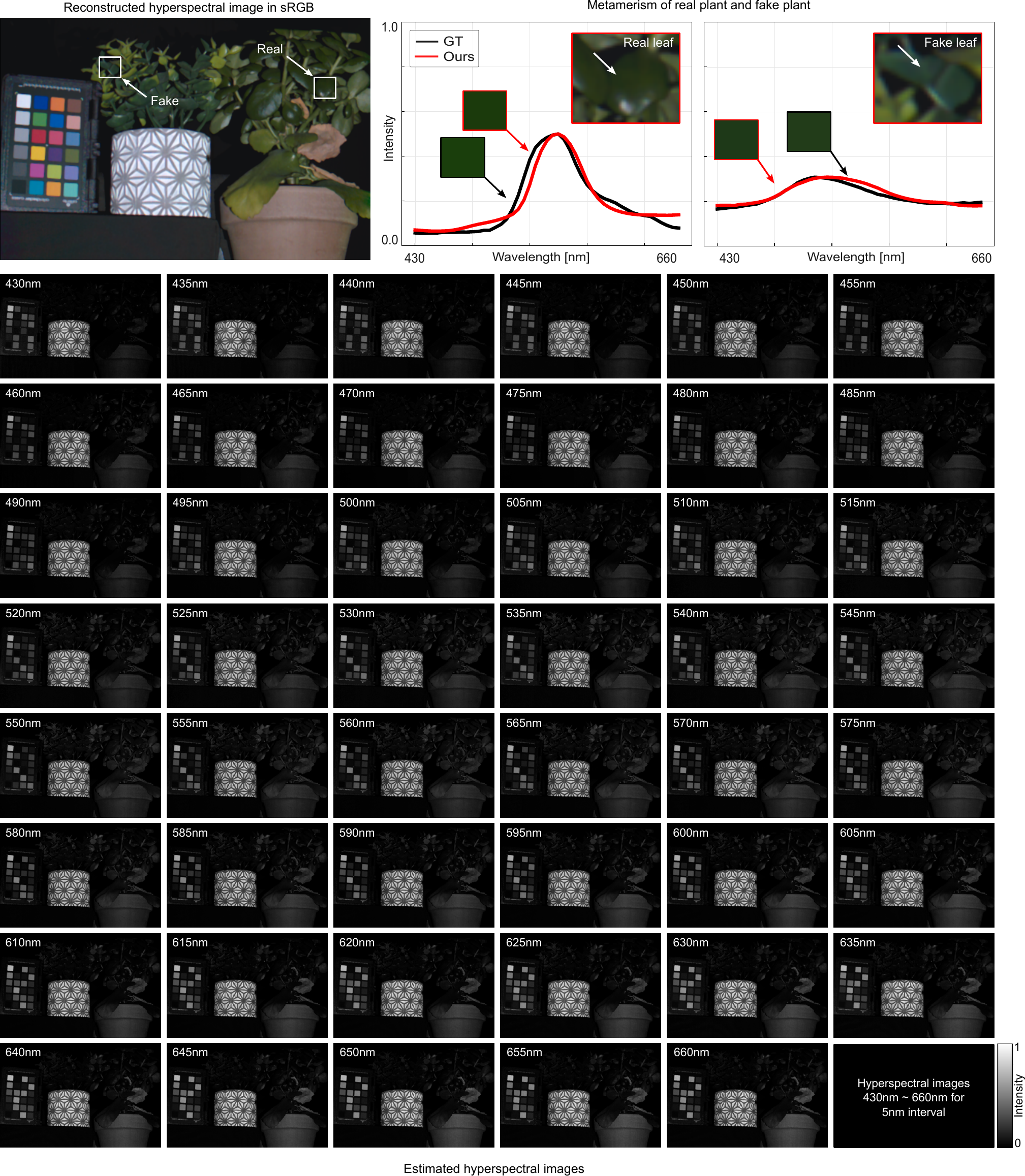} 
    \caption{\textbf{Hyperspectral Imaging.} We show the reconstructed hyperspectral image in sRGB, the metamerism of real plant and fake plant earned by spectroradiometer and the detailed hyperspectral image for 5 nm interval.}
    \label{fig:hyperspectral_scene3}
    \vfill 
\end{figure*}

\begin{figure*}[ht]
    \centering
    \vfill
    \includegraphics[height=0.9\textheight]{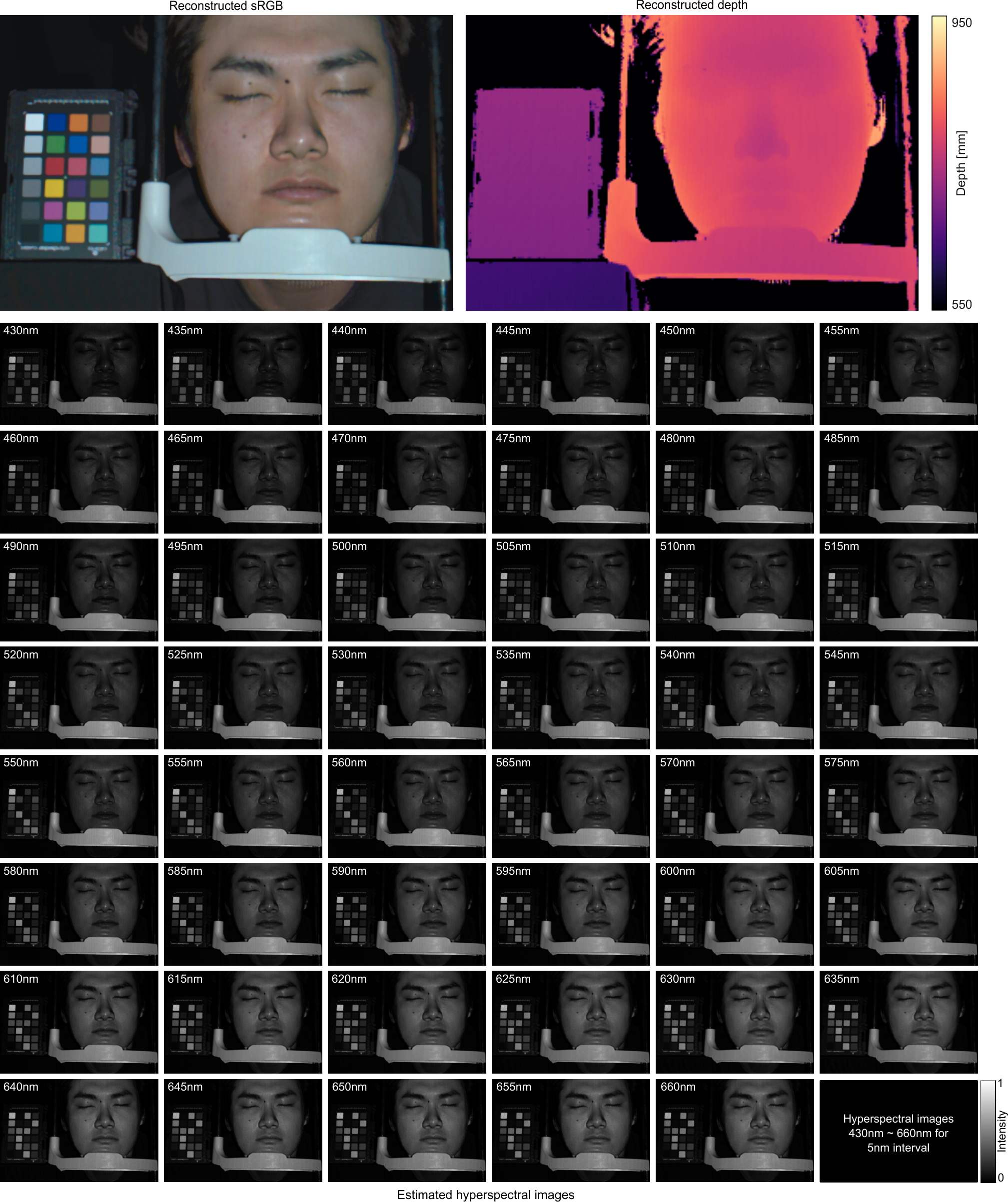} 
    \caption{\textbf{Hyperspectral 3D imaging.} We show the result of a face scanned reconstructed hyperspectral image in sRGB, reconstructed depth and the detailed hyperspectral image for 5 nm interval.}
    \label{fig:hyperspectral_scene5}
    \vfill 
\end{figure*}

We present an accurate and detailed result of reconstructed hyperspectral image in sRGB and hyperspectral images from 430\,nm to 660\,nm at 5\,nm intervals. To validate the fidelity of these reconstructions, we conduct a comparison with ground truth measurements obtained from a spectroradiometer. This comparison focuses on the spectral curves at specific points for each scenes shown in Figure~\ref{fig:hyperspectral_scene1}, ~\ref{fig:hyperspectral_scene2} and ~\ref{fig:hyperspectral_scene3}. We also show an additional result of face scanned result in ~\ref{fig:hyperspectral_scene5} showing 430\,nm to 660\,nm in 5\,nm intervals.

\subsection{Reconstruction of High-frequency Spectral Curves}
\begin{figure*}[ht]
    \centering
    \vfill
    \includegraphics[height=0.95\textheight]{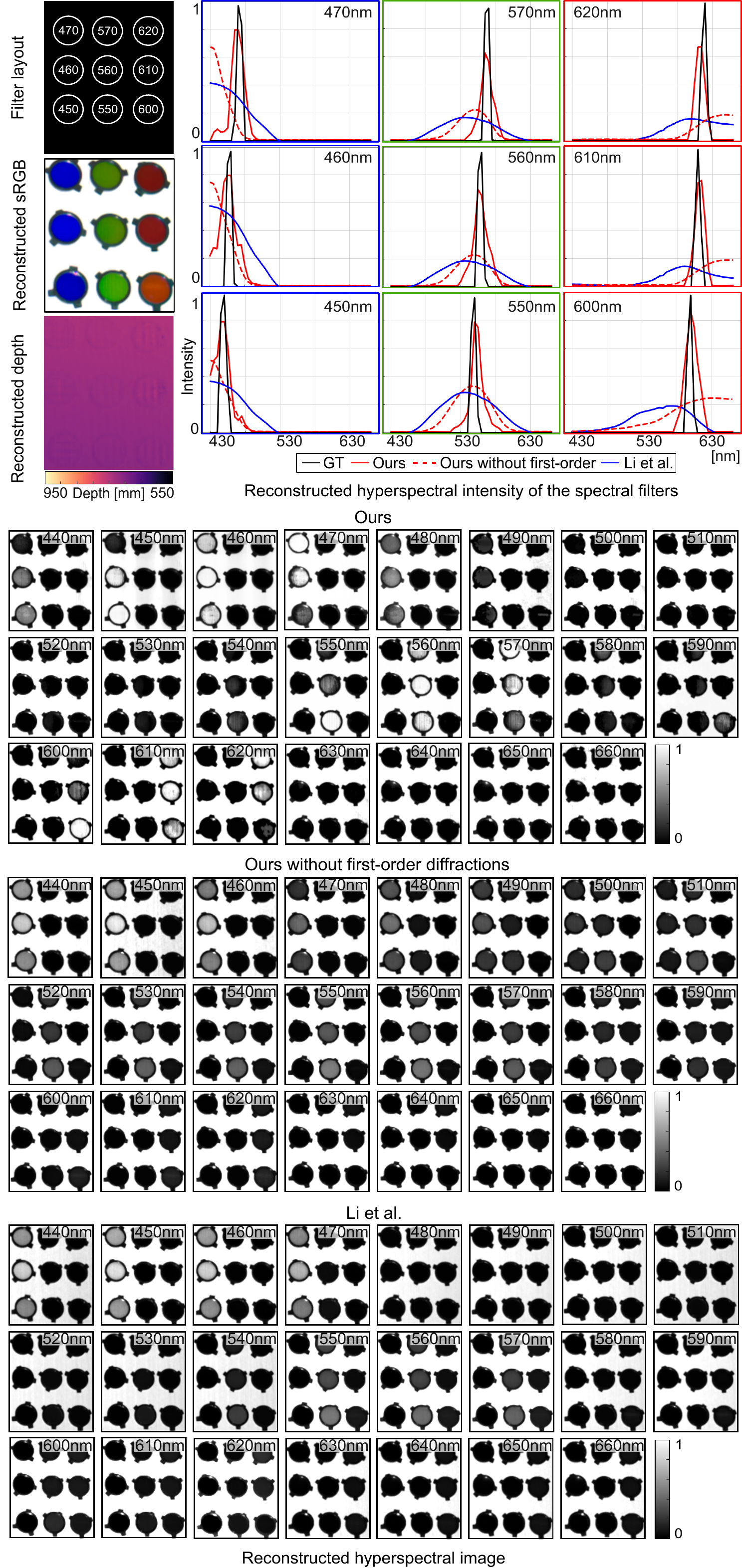} 
    \caption{\textbf{High-Frequency Evaluation.} We show the reconstructed hyperspectral images in sRGB and depth of high frequency band pass filter scene. We compare the reconstructed hyperspectral intensity and images with Li et al.\cite{li2019pro}.}
    \label{fig:hyperspectral_scene4}
    \vfill 
\end{figure*}

We compare the result of high-frequency spectral curve reconstruction of our DSL method with Li et al.~\cite{li2019pro}.  In Figure~\ref{fig:hyperspectral_scene4}, we show the reconstructed hyperspectral images generated by both methods, wavelengths from 440\,nm to 660\,nm at intervals of 10\,nm. Our DSL method accurately reconstructs the high-frequency hyperspectral spectral curves across each of the nine bandpass filters. In contrast, Li et al.~\cite{li2019pro} fails to reconstruct details of these curves. This comparison underscores the advanced capabilities of our DSL method in handling high-frequency spectral reconstruction. 

\bibliographystyle{plain}
\bibliography{supple_reference}

\end{CJK}